\input harvmac
\tolerance=10000



%
\def\omit#1{}
\def\eql{~=~}

\def\coeff#1#2{\relax{\textstyle {#1 \over #2}}\displaystyle}
\def\half{{1 \over 2}}

\def\ntwo{\cN\!=\!2}

 \def\cM{{\cal M}}
\def\cN{{\cal N}} \def\cO{{\cal O}}
 
 \def\cS{{\cal S}}
\def\cT{{\cal T}} 
\def\cV{{\cal V}} 
\def\cX{{\cal X}} 

\def\bfone{\relax{\rm 1\kern-.35em 1}}

%
\def\us{\bf}

\input amssym
\def\ZZ{\Bbb{Z}}

\def\IP{\Bbb{P}}

\def\nihil#1{{\it #1}}
\def\eprt#1{{\tt #1}}
\def\nup#1({Nucl.\ Phys.\ $\us {B#1}$\ (}
\def\plt#1({Phys.\ Lett.\ $\us  {#1B}$\ (}
\def\plt#1({Phys.\ Lett.\ $\us  {#1B}$\ (}
\def\cmp#1({Comm.\ Math.\ Phys.\ $\us  {#1}$\ (}
\def\prp#1({Phys.\ Rep.\ $\us  {#1}$\ (}
\def\prl#1({Phys.\ Rev.\ Lett.\ $\us  {#1}$\ (}
\def\prv#1({Phys.\ Rev.\ $\us  {#1}$\ (}
\def\mpl#1({Mod.\ Phys.\ Let.\ $\us  {A#1}$\ (}
\def\ijmp#1({Int.\ J.\ Mod.\ Phys.\ $\us{A#1}$\ (}
\def\atmp#1({Adv.\ Theor.\ Math.\ Phys.\ $\bf {#1}$\ (}
\def\cqg#1({Class.\ Quant.\ Grav.\ $\bf {#1}$\ (}
\def\jag#1({Jour.\ Alg.\ Geom.\ $\us {#1}$\ (}
\def\jhep#1({JHEP $\bf {#1}$\ (}

%


%



\def\BZ{\Bbb{Z}}
\def\BR{\Bbb{R}}
\def\BC{\Bbb{C}}
\def\BP{\Bbb{P}}

\def\CM{\cM}
\def\CN{\cN}

\def\CS{\cS}

\def\CW{{\cal W}}

%
%
\lref\KPW{A.\ Khavaev, K.\ Pilch and N.P.\ Warner, \nihil{New
Vacua of Gauged  ${\cal N}=8$ Supergravity in Five Dimensions},
\plt{487}  (2000) 14;  \eprt{hep-th/9812035}.}
%
\lref\FGPWa{D.~Z. Freedman, S.~S. Gubser, K.~Pilch, and N.~P. Warner,
\nihil{Renormalization Group Flows from Holography---Supersymmetry
and a c-Theorem,}   \atmp{3} (1999) 363; \eprt{hep-th/9904017} }
%
\lref\RGLMJS{ R.~G.~Leigh and M.~J.~Strassler, \nihil{Exactly marginal
operators and duality in four-dimensional N=1 supersymmetric gauge theory,}
\nup{447} (1995)  95; \eprt{hep-th/9503121}.}
\lref\ConTrunc{ B.~de Wit and H.~Nicolai, \nihil{The Consistency of the $S^7$
Truncation in D = 11 Supergravity,}
\nup{281}  (1987) 211.}
\lref\KPNWa{K.\ Pilch and N.P.\ Warner, \nihil{A New Supersymmetric
Compactification of Chiral IIB Supergravity,} \plt{487} (2000) 22;
\eprt{hep-th/0002192}.}
%
\lref\KPNWb{K.~Pilch and N.P.~Warner, \nihil{$\cN=2$ Supersymmetric
RG Flows and the IIB  Dilaton,} \nup{594} (2001) 209;
\eprt{hep-th/0004063}.}
%
\lref\KPNWc{
K.~Pilch and N.~P.~Warner,
\nihil{N=1 Supersymmetric Renormalization Group Flows from IIB Supergravity,}
\atmp{4} (2000) 627-677, \eprt{hep-th/0006066}.}
%
\lref\KPNWntwo{K.~Pilch and N.P.~Warner, \nihil{$\ntwo$ Supersymmetric
RG Flows and the IIB  Dilaton,} \nup{594} (2001) 209;
\eprt{hep-th/0004063}.}
%
\lref\LJR{ L.~J.~Romans, \nihil{New Compactifications of Chiral N=2,
 D = 10 Supergravity,} \plt{153} (1985)   392. }
%
\lref\IKEW{I.~R.~Klebanov and E.~Witten, \nihil{Superconformal Field Theory
on Threebranes at a Calabi-Yau  Singularity,} \nup{536} (1998)  199;
\eprt{hep-th/9807080}.}
%
\lref\IKMS{
I.~R.~Klebanov and M.~J.~Strassler,
\nihil{Supergravity and a confining gauge theory: Duality cascades and  
$\chi$-{SB}-resolution of naked singularities,} JHEP {\bf 0008} (2000)  052; 
\eprt{hep-th/0007191}.}
\lref\FGPWb{
D.~Z.~Freedman, S.~S.~Gubser, K.~Pilch and N.~P.~Warner,
JHEP {\bf 0007} (2000) 038, \eprt{hep-th/9906194}.}
\lref\GukovKK{
S.~Gukov,
\nihil{Comments on N = 2 AdS orbifolds},
Phys.\ Lett.\ B {\bf 439},  (1998) 23
\eprt{hep-th/9806180}.
}
\lref\DouglasSW{
M.~R.~Douglas and G.~W.~Moore,
\nihil{D-branes, Quivers, and {ALE} Instantons},
\eprt{hep-th/9603167}.
}
\lref\JohnsonPY{
C.~V.~Johnson and R.~C.~Myers,
\nihil{Aspects of type {IIB} theory on {ALE} spaces},
Phys.\ Rev.\ D {\bf 55}, 6382 (1997)
\eprt{hep-th/9610140}.
}
\lref\KachruYS{
S.~Kachru and E.~Silverstein,
\nihil{4d conformal theories and strings on orbifolds},
Phys.\ Rev.\ Lett.\  {\bf 80}, 4855 (1998)
\eprt{hep-th/9802183}.
}
\lref\DallAgataVB{
G.~Dall'Agata, C.~Herrmann and M.~Zagermann,
\nihil{General matter coupled $N = 4$ gauged supergravity in five
dimensions},
Nucl.\ Phys.\ B {\bf 612}, 123 (2001)
\eprt{hep-th/0103106}.
}
\lref\GunaydinFK{ M.~G\"{u}naydin and N.~Marcus,
 \nihil{The Spectrum Of The $S^5$ Compactification of The Chiral N=2, D = 10 Supergravity And The Unitary Supermultiplets 
of $U(2, 2/4)$}, Class.\ Quant.\ Grav.\  {\bf 2}, L11 (1985).
}

\lref\GunaydinUM{ M.~G\"{u}naydin, \nihil{Singleton And Doubleton
Supermultiplets Of  
Space-Time Supergroups And Infinite Spin Superalgebras }, CERN-TH-5500/89 {\it 
Proceedings of the  Conf. on Supermembranes and Physics in
(2+1)-Dimensions, Trieste,  
Italy, Jul 17-21, 1989}, eds. M. Duff, C. Pope and E. Sezgin ( World
Scientific, 1990).   
}
\lref\KimEZ{ H.~J.~Kim, L.~J.~Romans and P.~van Nieuwenhuizen,
\nihil{The Mass Spectrum of Chiral N=2 D = 10 Supergravity On $S^5$},
Phys.\ Rev.\ D {\bf 32}, 389 (1985). 
}
\lref\GZ{M. G\"{u}naydin and M. Zagermann,
\nihil{The Gauging of Five-dimensional, N=2 Maxwell-Einstein Supergravity 
Theories coupled to Tensor
      Multiplets},
Nucl.\ Phys.\ B {\bf 572}, 131 (2000)
\eprt{hep-th/9912027}.
}
\lref\BHLT{K. Behrndt, C. Herrmann, J. Louis and  S. Thomas,
\nihil{Domain walls in five dimensional supergravity with 
non-trivial hypermultiplets},
JHEP\ {\bf 0101}, 011 (2001)
\eprt{hep-th/0008112}.
}
\lref\CDKV{A. Ceresole, G. Dall'Agata, R. Kallosh and A. Van Proeyen,
\nihil{Hypermultiplets, domain walls and supersymmetric attractors},
Phys.\ Rev.\  {\bf D64}, 104006 (2001)
\eprt{hep-th/0104056}.
}
\lref\TPvN{
P.K. Townsend, K. Pilch and  P. van Nieuwenhuizen,
\nihil{Selfduality in odd dimensions},
Phys.\ Lett.\ {\bf 136B}, 38 (1984), Addendum-ibid.{\bf 137B}, 443 (1984).
}
\lref\IntriligatorIG{
K.~A.~Intriligator,
\nihil{Bonus symmetries of $\CN = 4$ super-Yang-Mills correlation
functions via  {AdS} duality},
Nucl.\ Phys.\ B {\bf 551}, 575 (1999)
\eprt{hep-th/9811047}.
}
\lref\GubserIA{
S.~Gubser, N.~Nekrasov and S.~Shatashvili,
\nihil{Generalized conifolds and four dimensional $N = 1$ superconformal
theories},
JHEP {\bf 9905}, 003 (1999)
\eprt{hep-th/9811230}.
}
\lref\BergmanQI{
A.~Bergman and C.~P.~Herzog,
\nihil{The volume of some non-spherical horizons and the {AdS/CFT}
correspondence},
\eprt{hep-th/0108020}.
}
\lref\OzHR{
Y.~Oz and J.~Terning,
\nihil{Orbifolds of {AdS$_5\times S^5$} and 4d conformal field theories,}
Nucl.\ Phys.\ B {\bf 532}, 163 (1998)
\eprt{hep-th/9803167}.
}
\lref\FerraraUR{
S.~Ferrara and A.~Zaffaroni,
\nihil{$N = 1,2$ {4D} superconformal field theories and supergravity in
{AdS$_5$},} 
Phys.\ Lett.\ B {\bf 431}, 49 (1998)
\eprt{hep-th/9803060}.
}
\lref\LawrenceJA{
A.~E.~Lawrence, N.~Nekrasov and C.~Vafa,
\nihil{On conformal field theories in four dimensions,}
Nucl.\ Phys.\ B {\bf 533}, 199 (1998)
\eprt{hep-th/9803015}.
}
\lref\BerensteinHY{
D.~Berenstein and R.~G.~Leigh,
\nihil{Discrete torsion, {AdS/CFT} and duality},
JHEP {\bf 0001}, 038 (2000)
\eprt{hep-th/0001055}.
}
\lref\VafaIH{
C.~Vafa,
\nihil{Quantum Symmetries Of String Vacua,}
Mod.\ Phys.\ Lett.\ A {\bf 4}, 1615 (1989).
}
\lref\WittenKZ{
E.~Witten,
\nihil{New ``gauge'' theories in six dimensions,}
JHEP {\bf 9801}, 001 (1998),
Adv.\ Theor.\ Math.\ Phys.\  {\bf 2}, 61 (1998)
\eprt{hep-th/9710065}.
}
\lref\BerensteinDW{
D.~Berenstein, R.~Corrado and J.~Distler,
\nihil{Aspects of {ALE} matrix models and twisted matrix strings,}
Phys.\ Rev.\ D {\bf 58}, 026005 (1998)
\eprt{hep-th/9712049}.
}
\lref\GRW{M.\ G\"unaydin, L.J.\ Romans and N.P.\ Warner,
\nihil{Gauged $N=8$ Supergravity in Five Dimensions,}
Phys.~Lett.~{\bf 154B} (1985) 268; \nihil{Compact and Non-Compact 
Gauged Supergravity Theories in Five Dimensions,} 
\nup{272} (1986) 598.
}
\lref\PerniciJU{M.~Pernici, K.~Pilch and P.~van Nieuwenhuizen,
\nihil{Gauged $N=8$ $D = 5$ Supergravity,}
Nucl.\ Phys.\ B {\bf 259}, 460 (1985).
}
\lref\AwadaEP{
M.~Awada and P.~K.~Townsend,
\nihil{$N=4$ {Maxwell-Einstein} Supergravity In Five Dimensions And
Its $SU(2)$ Gauging}, 
Nucl.\ Phys.\ B {\bf 255}, 617 (1985).
}
\lref\RomansPS{
L.~J.~Romans,
\nihil{Gauged {$N=4$} Supergravities In Five Dimensions And Their
Magnetovac Backgrounds,} 
Nucl.\ Phys.\ B {\bf 267}, 433 (1986).
}
\lref\HenningsonGX{M.~Henningson and K.~Skenderis, {\it The
holographic Weyl anomaly,} JHEP {\bf 9807}, 023 (1998)
\eprt{hep-th/9806087}.
}
\lref\PerniciXX{M.~Pernici, K.~Pilch and P.~van Nieuwenhuizen, {\it
Gauged Maximally Extended Supergravity In Seven-Dimensions,} Phys.\
Lett.\ B {\bf 143}, 103 (1984).
}
%
\lref\GubserVD{
S.~S.~Gubser,
\nihil{Einstein manifolds and conformal field theories,}
Phys.\ Rev.\ D {\bf 59}, 025006 (1999)
\eprt{hep-th/9807164}.
}%
\lref\LopezZF{
E.~Lopez,
\nihil{A family of $N = 1$ $SU(N)^k$ theories from branes at
singularities},
JHEP {\bf 9902}, 019 (1999)
\eprt{hep-th/9812025}.
}
\lref\WittenSC{
E.~Witten,
\nihil{Solutions of four-dimensional field theories via M-theory,}
Nucl.\ Phys.\ B {\bf 500}, 3 (1997)
\eprt{hep-th/9703166}.
}
\lref\DoreyQJ{
N.~Dorey, T.~J.~Hollowood and S.~Prem Kumar,
\nihil{An exact elliptic superpotential for $N = 1^*$ deformations of
finite  $N = 2$  gauge theories,}
\eprt{hep-th/0108221}.
}
\lref\IntriligatorIG{
K.~A.~Intriligator,
\nihil{Bonus symmetries of $N = 4$ super-Yang-Mills correlation functions via  AdS duality,}
Nucl.\ Phys.\ B {\bf 551}, 575 (1999)
\eprt{hep-th/9811047}.
}
\lref\KhavaevYG{
A.~Khavaev and N.~P.~Warner,
\nihil{An N = 1 supersymmetric Coulomb flow in IIB supergravity,}
Phys.\ Lett.\ B {\bf 522}, 181 (2001)
\eprt{hep-th/0106032}.
}
\lref\ConTrunc{
B.\ de Wit and H.\ Nicolai, \nihil{On the Relation
Between $d=4$ and $d=11$ Supergravity,} Nucl.~Phys.~{\bf B243}
(1984) 91; \hfil \break
B.\ de Wit, H.\ Nicolai and N.P.\ Warner,
\nihil{The Embedding of Gauged $N=8$ Supergravity into $d=11$
Supergravity,} Nucl.~Phys.~{\bf B255} (1985) 29; \hfil \break
B.\ de Wit, H.\ Nicolai, \nihil{The Consistency of the
$S^7$  Truncation in $d = 11$ Supergravity,} Nucl.~Phys.~{\bf B281}
(1987) 211. \hfill \break
H. \ Nastase, D. \ Vaman and P.\ van Nieuwenhuizen, \nihil{Consistent 
nonlinear KK reduction of 11d supergravity on $AdS_7\times S_4$ and 
self-duality in odd dimensions,}  Nucl.\ Phys.\ B {\bf 581}, 179 (2000), 
\eprt{hep-th/9905075}.
}
\lref\KarchPV{
A.~Karch, D.~Lust and A.~Miemiec,
\nihil{New $N = 1$ superconformal field theories and their supergravity  
description,}
Phys.\ Lett.\ B {\bf 454}, 265 (1999)
\eprt{hep-th/9901041}.
}%
\lref\MorrisonCS{
D.~R.~Morrison and M.~R.~Plesser,
\nihil{Non-spherical horizons. I,}
Adv.\ Theor.\ Math.\ Phys.\  {\bf 3}, 1 (1999)
\eprt{hep-th/9810201}.
}%
\Title{ \vbox{ \hbox{CITUSC/02-007, USC-02/01} 
\hbox{\tt hep-th/0203057} }} {\vbox{\vskip -1.0cm
\centerline{\hbox
{Orbifolds and Flows from Gauged Supergravity}}
\vskip 8 pt
\centerline{
\hbox{}}}}
\vskip -0.9cm
\centerline{Richard Corrado$^1$, Murat G\"unaydin$^2$, Nicholas P.\
Warner$^1$, and Marco Zagermann$^3$ } 
\bigskip
\centerline{$^1${\it Department of Physics and Astronomy}} 
\centerline{{\it and}}
\centerline{{\it CIT-USC Center for
Theoretical Physics}}
\centerline{{\it University of Southern California}} 
\centerline{{\it Los Angeles, CA
90089-0484, USA}} 
\medskip
\centerline{$^2${\it Physics Department}} 
\centerline{{\it Pennsylvania State University}}
\centerline{{\it University Park, PA 16802, USA}}
\medskip
\centerline{$^3${\it Fachbereich Physik}} 
\centerline{{\it Martin-Luther-Universit\"{a}t 
Halle-Wittenberg,}}
\centerline{{\it Friedemann-Bach-Platz 6, D-06099 Halle, Germany}}
\bigskip

\centerline{{Abstract}}
\medskip
We examine orbifolds of the IIB string via gauged supergravity. For the
gravity duals of the $A_{n-1}$ quiver gauge theories, 
we extract the massless degrees of
freedom and assemble them into multiplets of $\CN=4$ gauged
supergravity in five dimensions. We examine the embedding of the gauge
group into the isometry group of the scalar manifold, as well as
the symmetries of the scalar potential. From this we 
find that there is a large $SU(1,n)$ symmetry group which relates
different RG flows in the dual quiver gauge theory. We find that this
symmetry implies an extension of the usual duality between
ten-dimensional IIB solutions which involves exchanging geometric
moduli with background fluxes. 

\Date{\sl {March, 2002}}


\parskip=4pt plus 15pt minus 1pt
\baselineskip=15pt plus 2pt minus 1pt

\newsec{Introduction}

Gauged supergravity has proven 
to be a remarkably effective tool in the construction and study of
holographic RG flows.  In this general approach,  one uses a  
five-dimensional supergravity theory to capture and simplify the
details of what is  usually a far more complicated ten-dimensional
supergravity theory.  This has been most extensively employed in the
study of flows of $\cN=4$ supersymmetric Yang-Mills theory  and their
holographic duals in IIB supergravity. In this instance, 
gauged $\cN=8$  supergravity in five dimensions \refs{\GRW,\PerniciJU}
captures (as a consistent  
truncation) essentially all the perturbations of $\cN=4$ Yang-Mills
that involve gauge  
invariant bilinear chiral operators.  There has also been work on
using  more general  
five-dimensional $\cN=2$ gauged supergravity theories to examine flow
solutions (see, for example,  \refs{\BHLT,\CDKV}) and, while this is 
interesting,  there is often an unresolved   
problem in determining  the dual theory on the  brane and then
establishing the duality  
precisely.    This is not an issue  for gauged $\cN=8$ supergravity
precisely because of  
its well-established connection with the $S^5$ compactification of IIB
supergravity  
\refs{\GunaydinFK,\KimEZ}  and hence with  $\cN=4$  supersymmetric
Yang-Mills on  
$D3$-branes.

In this paper we will examine flow solutions in five-dimensional,
$\cN=4$ gauged supergravity theories 
\refs{\AwadaEP,\RomansPS,\DallAgataVB}.  Our purpose is to study 
perturbations of, and flows in, the $\cN=2$ supersymmetric quiver gauge 
theories in four dimensions.   The quiver theories are, of course,
conformal and satisfy the relationship, $c=a$,
of central charges that is essential to a holographic theory
\HenningsonGX.   
These theories are related to the $\cN=4$ theory via their
construction as the world-volume theory on D3-branes at an orbifold
singularity~\refs{\DouglasSW,\JohnsonPY}.  It is thus straightforward
to identify the holographic duals of quiver theories in terms of the
IIB superstring on a background of the form $AdS_5 \times
S^5/\Gamma$~\KachruYS, where $\Gamma$ is 
an  appropriately chosen discrete subgroup of 
$SU(2)_L \subset SU(2)_L  \times SU(2)_R
\times U(1) \subset SO(6)$, and where $SO(6)$ is the isometry group
of $S^5$.

Our first task will be to identify the  correct gauging and matter 
content for the $\cN=4$ gauged supergravity theory so as to obtain 
a subsector of the holographic dual of the quiver theories.  We will
restrict our attention to the $A_{n-1}$ quivers that are obtained by
taking $\ZZ_n$ orbifolds.  From the supergravity perspective, the
problem is to find the proper number of massless vector and tensor
multiplets, along with their charges.   This problem will be resolved
here in exactly the same manner that it was in the 1980's when 
similar ambiguities had to be resolved in the proper gauging
of maximal supergravities (see, for example, \PerniciXX):  We will
study the linearization of the corresponding compactification of the 
IIB theory, and this will give us precisely the proper field
content and charges. 

An $\cN=2$ superconformal Yang-Mills theory has an $SU(2)_R \times
U(1)$ R-symmetry,  and so this must be the gauge group of the
supergravity.   
One can then get further  insight into the gauging by looking at the
untwisted sector of  
the orbifold. We will examine this in greater detail in section~3, but
here we note that  
the untwisted sector must include the $SU(2)_L$ invariant sector of the $S^5$ 
compactification of the IIB theory.  In truncating to $\cN=8$ gauged
supergravity we see  
that this $SU(2)_L$ invariant sector must reduce to the $SU(2)$
invariant sector  
considered in \KPW. This fact was noted in \KPW\ and indeed was part
of the motivation   
for the use of the truncation to $SU(2)$ singlets.   The result is
$\cN =4$ gauged 
supergravity coupled to two tensor multiplets, and the gauge group is,
of course,   
$SU(2)_R \times U(1)$.  In \KPW\ the scalar coset was shown to be  
\eqn\utcoset{{ SO(5,2) \over SO(5)\times SO(2)}\times SO(1,1).}
The group $SO(5,2)$ contains an obvious $SO(3) \times SO(2) \times SO(2)$ and 
the $SU(2)_R$ gauge symmetry of the supergravity is to be identified
with the $SO(3)$, 
while the $U(1)$ gauge symmetry is the diagonal subgroup of  $SO(2)
\times SO(2)$.

It is also worth remembering that this $SU(2)_R \times U(1)$ commutes
with an $SU(1,1)$  
in $SO(5,2)$, and that this $SU(1,1)$ is naturally identified with the
$SU(1,1)$ of the  
dilaton and axion in the original IIB theory.  The supergravity
potential is invariant  
under the action of this $SU(1,1)$, and since the dilaton and axion
are dual to the  
gauge coupling and $\theta$-angle, this invariance reflects the fact
that the complex  
gauge coupling is a freely chooseable parameter at the UV fixed point.

We will show in section~4 that the foregoing generalizes very naturally to
the quiver theories.  First, we will show that the supergravity theory
we seek is $\cN=4$ gauged supergravity coupled to $2n$ tensor
multiplets and either  one or three  
vector multiplets.  The corresponding scalar manifold is of the form
\eqn\scalarM{\cS ~=~ {SO(5,2n+q) \over 
SO(5) \times SO(2n+q)}  \times SO(1,1)\,,}
where $2n$ is the number of tensor multiplets and $q$ is the number of
vector multiplets.  The $SO(1,1)$ factor is parameterized by the
``dilaton'' in the five-dimensional gravity supermultiplet. 
There is an obvious 
$SO(3) \times (SO(2))^{n+1}$ 
subgroup of $SO(5,2n)$, and $SU(2)_R$ is still to be identified with
the $SO(3)$ and the  
$U(1)$ is the diagonal embedding in the product of $SO(2)$'s.   The
supergravity gauge  
symmetry, $SU(2)_R \times U(1)$, thus commutes with $SU(1,n)$ in
$SO(5,2n)$.  We will  
show in section~4  that this $SU(1,n)$ is an invariance of the
potential in the gauged  
$\cN=4$ theory.  Moreover we will argue that the scalars that
parameterize the coset 
\eqn\CoupCoset{\cT ~=~ {SU(1,n) \over U(1)\times SU(n) } \,, }
are in fact the duals of the $n$ distinct complex coupling constants 
of the $n$ distinct $SU(N)$ gauge group factors 
associated with the nodes of the quiver diagram.  The
$SU(1,n)$  
invariance thus represents the fact that these couplings are freely 
chooseable parameters.

Having identified the correct matter content and gauging
we will go on in section~5 to analyze part of the
corresponding supergravity potential.  We first describe how to
reconstruct the results of \KPW, and we then describe how  
the flows of \refs{\IKEW,\GubserIA} must fit within the $\cN =4$
gauged supergravity.  Having done this, we discover a rather
surprising result:  there is an $SU(n)$ symmetry that must map all
of the flows of \refs{\IKEW,\GubserIA} onto the flow of \FGPWa.  To be
precise, 
the $SU(n) \subset SU(1,n)$ is a symmetry of the supergravity
action that maps the supergravity scalars that describe the flow
of \FGPWa\ onto any and all of the flows of  \refs{\IKEW,\GubserIA}.
In particular, we find that the critical point of \FGPWa\ is extended
to a complex $(n-1)$-dimensional surface in the $\cN=4$ supergravity
theory corresponding to the $A_{n-1}$ quiver theory.  Indeed, if $g_i$
and $m_j$ are the gauge coupling constants and masses associated with
the $i^{\rm th}$ node of the quiver, then the critical surface
is a $\IP^{n-1}$ parameterized by homogeneous coordinates
$ g_i^2/m_i$.  In particular the flows of \refs{\IKEW, \GubserIA} and 
\FGPWa\ are represented by distinct points on this surface.

This is a rather remarkable claim.  As we will see, it is almost
trivial from the  
five-dimensional perspective, but from the ten-dimensional IIB
perspective, it means  
that there must be a continuous symmetry that trades the tensor gauge
field fluxes  
obtained in \KPNWc\ on a topologically trivial manifold for K\"ahler
moduli of blow-ups  
on the resolved orbifold\foot{More precisely, these are complex structure
 moduli of the deformation of the orbifold singularity in the 3-fold
$\BC^2/\BZ_n\times  
\BC$.}. Such a symmetry is not unprecedented:  It is certainly not the
first time that one has discovered that a very simple, manifest
symmetry in low dimensions could have  
remarkable, and unexpected geometric consequences in higher
dimensions. 
In this case, the IIB string theory enjoys a discrete duality symmetry
which acts on the metric moduli and fluxes and it is promoted to a
continuous symmetry of the supergravity sector. We shall have more to
say about this in our discussion in section~6.

There were also very
indirect hints that  there might be such a symmetry.    
The flows are similar in their field theory   
description, and have the same symmetry and supersymmetry.  
Moreover all these flows 
give rise to infra-red  fixed points with the same central 
charge\foot{In making this 
statement we are thinking of the flow of \FGPWa\ as being 
embedded in the $\cN=2$ quiver 
theory and not in the original setting of $\cN=4$ Yang-Mills theory, 
and so the flows 
start with the same value of $c_{UV}$.}: ${c_{IR} \over c_{UV}} = 
{27 \over 32}$. For 
the resolved conifolds, this result about the central charges was 
established for $n=2$ 
in \GubserVD, and rather more recently for general $n$ in \BergmanQI.  
For the flows 
involving only fluxes on the unresolved $S^5/\ZZ_n$, the ratio of 
central charges 
follows from the results of \KPW\ and the observation that this 
particular flow lies 
entirely in the untwisted sector.    Our results here will show, 
amongst other things, 
how this generalizes to a continuous family of flows that spans 
smoothly across all 
twisted sectors.

Since we are using a five-dimensional theory to describe a 
solution that lives in IIB supergravity in ten-dimensions, we are once
again haunted by an old ghost of gauged supergravity: Consistent
Truncation.  The issue is whether a solution of the five-dimensional
theory  really does ``lift'' to an exact solution in ten dimensions.  The fact
that one  can do this is  fairly well established for the maximal gauged
supergravity theories in four and seven dimensions \ConTrunc.
While there is no formal proof in five dimensions, the result
is very plausible, and many lifts   have been explicitly constructed  
(see, for example,  \refs{\KPNWa, \KPNWb, \KPNWc, \KhavaevYG}). 
However, our work here goes beyond the ``established'' results of 
consistent truncation:
we are using half-maximal supersymmetry, with added tensor multiplets.
There is thus a legitimate concern that we may be working with only an
{\it effective}  five-dimensional  theory, and our results may not have
exact ten-dimensional lifts.    Based on our experience of consistent 
truncation we know that if it fails then one typically finds a solution
in lower dimensions that has no analog in higher dimensions, or vice
versa. 
Moreover, there may be a symmetry mismatch.
In the instance we consider here, we have a family of solutions in both
five and ten dimensions, with the same symmetry and supersymmetry, and
the families are generated from flows with the same initial data (for
which 
there {\it is} an exact perturbative correspondence).  Moreover,  we
will see that 
one can understand the result from the field theory on the brane, and
the recent work in \BergmanQI\ shows that the ten-dimensional
solutions have the correct cosmological constant, or central charge for the
brane theory.
We therefore feel that we are on very solid ground, and furthermore
the results presented here strongly suggest
that the half-maximal gauged supergravities also
represent a class of consistent  truncations of the IIB theory.

\newsec{A Class of RG Flows in $\CN=2$ Quiver Gauge Theories}

We will be primarily concerned with D3-branes at $A_{n-1}$ singularities
$\BC^2/\BZ_n$. If the pullback of the orbifold projection to the brane
world-volume is taken to be the regular representation, the gauge
theory on the $N$ branes is a four-dimensional $\CN=2$ $U(N)^n$ gauge 
theory with $n$ adjoint vector multiplets and $n$ hypermultiplets 
transforming in
the bifundamental representations of adjacent groups~\DouglasSW. This
field content can be associated to the extended Dynkin diagram of
$A_{n-1}$ (called the ``quiver''). 
As is usual in brane world-volume theories,  the scalar degrees
of freedom parameterize the space transverse to the branes. Here
the scalars in the hypermultiplets describe the $\BC^2/\BZ_2$
directions, while those in the vector multiplets describe the
remaining transverse $\BR^2=\BC$. 

We can represent the $\cN=2$ supersymmetric hypermultiplets in 
terms of $\cN=1$ chiral fields
$(A_i,B_i)$ where the $A_i$ are in the $({\bf N}_{i-1},\bar{\bf
N}_{i})$ representation of $U(N_{i-1})\times U(N_i)$ and the $B_i$ are
in the $(\bar{\bf N}_{i-1}, {\bf N}_i)$. The $\cN=1$ chiral fields in
the $\CN=2$ vector multiplets will be denoted by $\Phi_i$ and are in
the adjoints of $U(N_i)$. The $\cN=1$ superpotential is then
\eqn\GTsuppot{W= \sum_i ~\lambda_i~{\rm Tr}~\int d^2\theta \left( 
B_{i+1} \Phi_i A_{i+1} -A_i \Phi_i B_i \right). }
For $\cN=2$ supersymmetry, the Yukawa couplings, $\lambda_j$, have to
be equal to the  
gauge couplings $g_j=\sqrt{ 4\pi/{\rm Im}~\tau_j}$ associated with
each node of the quiver theory. 
In the orbifold  
gauge theory\foot{Specifically, we mean the point in the string moduli
space where the  
worldsheet orbifold CFT is valid.}, all these gauge couplings are
equal, $g_j = g$.   
(This is because they are all equal to $\sqrt{n}$ times the gauge
coupling of the  
original $\cN=4$ Yang-Mills theory before the orbifoldization.) These
$\cN=2$ theories  
are superconformal, and have the R-symmetry group $SU(2)_R\times U(1)$. 
Under the Cartan generator $J_3$ of $SU(2)_R$, the hypermultiplets 
$(A_i,B_i)$ have charge $J_3=1$, while the 
$\Phi_i$ are neutral.  Under the generator, $Y$, of the $U(1)$ factor,
the fields  
$\Phi_i$  have charge $Y=2$, while the $(A_i,B_i)$ are neutral. The
superconformal,  
$\cN=1$, R-charge is given by the expression
\eqn\Rcharge{R_{sc} = {Y + 2\, J_3 \over 3} .}
The mass dimension of a field with this R-charge is given by 
$d = 3 R_{sc}/2 + \gamma$, where $\gamma$ is the anomalous dimension.

In principle one can also add D and F-terms for the $U(1)$ factors. In
the usual 
brane-probe analysis, these terms allow the resolution of the orbifold
singularity in the moduli space of the probe.   In the infrared these
$U(1)$'s decouple and so we cannot use D and F-terms to resolve the
singularity in the solution of the IIB theory, and so we
must look at relevant deformations~\KachruYS. 

Klebanov and Witten~\IKEW\foot{The definition of $(A_2,B_2)$
in~\IKEW\ is opposite to ours. Here $A_2=B_2^{KW}$, $B_2=A_2^{KW}$.} 
considered a deformation
to an $\cN=1$ theory by a twisted mass term for the adjoint chiral fields
\eqn\Aoneopdef{
\widetilde{\CW_2} = {m \over 2}~ {\rm Tr}~\int d^2\theta \, (\Phi_1^2 - 
\Phi_2^2).}
Geometrically, this deformation corresponds to a complex
structure deformation of the manifold $\BC^2/\BZ_2 \times \BC$ and
removes the orbifold singularity, leaving a conical singularity at the
origin. This deformation leads to a theory in the infra-red that can be
studied by integrating out the massive $\Phi_{1,2}$ multiplets. 
Doing this in~\GTsuppot\
leads to a quartic superpotential for the 
chiral fields $A_i,B_i$,
\eqn\Aonequarsuppot{
\widetilde{W}_2= {g^2 \over m}~ {\rm Tr}~\int d^2\theta \, (
A_1 B_1 B_2 A_2 -A_1 A_2 B_2 B_1 ).}
In~\IKEW, it was explained that the $\CN=1$ theory in the IR is the
theory that describes D3-branes at the conifold singularity. 

Another obvious $\CN=1$ deformation of the $A_1$ quiver theory is by
the untwisted mass term:
\eqn\Aoneuntwist{
\CW_2 = {m \over 2}~ {\rm Tr}~\int d^2\theta \, 
(\Phi_1^2 + \Phi_2^2).}
This deformation is analogous to the deformation that 
gives a mass to one of the $\CN=1$ chiral fields in
the $\CN=4$ super-Yang--Mills multiplet~\refs{\RGLMJS,\KarchPV,\FGPWa}. 
Now we obtain an $\CN=1$  theory with quartic superpotential
\eqn\Aoneuntwistquarsuppot{
W_2= {g^2 \over m}~ {\rm Tr}~\int d^2\theta \, \left(
(A_1 B_1)(A_1B_1-B_2A_2)+  (A_2 B_2)(A_2 B_2-B_1A_1) \right).}
The geometry dual to the flow generated by~\Aoneuntwist\ can be
obtained as a $\BZ_2$ orbifold of the ten-dimensional lift~\KPNWc\
of the flow  found in \FGPWa. 

More generally, we can consider the   
deformations~\refs{\GubserIA,\LopezZF,\BergmanQI,\DoreyQJ}:
\eqn\Genopdef{
\CW_n
= \sum_i {m_i \over 2}~ {\rm Tr}~\int d^2\theta \, \Phi_i^2 .}
(By a change of basis, we could rewrite this in terms of $n-1$ twisted 
superfield 
operators $\Phi_i^2 - \Phi_{i+1}^2$ and the untwisted operator 
$\sum_{i=1}^n\Phi_i^2$, 
but we will not need its explicit form.) We note that in the IR, 
the overall mass scale 
of the {\it complex } parameters $m_i$ will decouple. For the
perturbation to be  
non-trivial, at least 
one of the $m_i$, say $m_j$, must be non-zero and can therefore 
be used to set the 
overall scale. The ratios $m_i/m_j$ yield inhomogeneous coordinates 
on $\BP^{n-1}$. This 
complex projective space is the 
moduli space of 
couplings of the deformations~\Genopdef. If all the $m_i$ are non-zero 
then we expect a 
class of flows, generalizing those of Leigh and Strassler, to a family 
of fixed points 
parameterized by $m_i/m_j$.   These fixed point theories will all have 
${c_{IR} \over c_{UV}} = {27 \over 32}$.   

If one, or more, of the
$m_i$ vanish then it is, {\it a priori}, less 
clear what will happen: there might not be a fixed point, or, if there
is, the central charge may be higher than that obtained when the
$m_i\neq 0$.
Thus we expect  
a surface of fixed points with ${c_{IR} \over c_{UV}} = {27 \over 32}$
described by $\BP^{n-1}$, with all
the hypersurfaces  $\BP^{n-2}$ defined by 
$m_i =0$ excised.  On these hypersurfaces, and their intersections, there 
are potentially different fixed-point theories with ${c_{IR} \over c_{UV}} 
\ge  {27 \over 32}$.  Our analysis below will more properly elucidate
the structure of the manifold of fixed points. 

It is relatively straightforward to use the methods of~\RGLMJS\
to see how the foregoing emerges from the conditions for conformal  
invariance. We begin with the quiver theory and its {\it a priori}
independent collection of gauge couplings, $\tau_i$, Yukawa couplings,
$\lambda_i$, and adjoint masses $m_i$. 
Whenever one of the $m_i$ is non-zero, we integrate 
out the corresponding $\Phi_i$\foot{Alternatively, we can leave the
field $\Phi_i$ in the theory. Then the vanishing of the beta function
for $m_i$ fixes $\gamma_{\Phi_i}=\half$ for all $i$ such that $m_i\neq
0$. Applying this to the beta functions for the Yukawa couplings will
leave us with the equations~(2.10).}. This leads to a superpotential
\eqn\partialsuppot{
\eqalign{&\sum_{i|m_i\neq 0} \, 
c_i~ {\rm Tr}~\int d^2\theta \, 
(A_{i+1}B_{i+1}-B_i A_i )^2 \cr
& +\sum_{i|m_i=0}~\lambda_i~ {\rm Tr}~\int d^2\theta \left( 
B_{i+1} \Phi_i A_{i+1} -A_i \Phi_i B_i\right), }}
where $c_i=\lambda_i^2 /(2m_i)$ and $i|m_i = 0$ denotes the
set of indices, $i$, such that $m_i =0 $.  Let the number of $\Phi_i$ 
that we have given mass to be $\mu$. Then
there are $\mu$ couplings $c_i$ and $n-\mu$ couplings $\lambda_i$
left. Vanishing of the
$\beta$-function for the gauge coupling requires that 
\eqn\betag{ 
\gamma_{A_i} + \gamma_{B_i}+\gamma_{A_{i+1}} + \gamma_{B_{i+1}}
+ \delta_{m_i,0}~(2\gamma_{\Phi_i} - 1) +1   = 0,~\forall~ i,}
where $\gamma_X$ is the anomalous dimension of the field $X$. The
vanishing of the $\beta_{\lambda_i}$ and the $\beta_{c_i}$ lead to the
equations 
\eqn\betacoup{\eqalign{
 2 \gamma_{\Phi_i} + \gamma_{A_i} + \gamma_{B_i} 
+\gamma_{A_{i+1}} + \gamma_{B_{i+1}} =0,~&~\forall~ i|m_i=0,  \cr
\gamma_{A_i} + \gamma_{B_i} 
+\gamma_{A_{i+1}} + \gamma_{B_{i+1}}+1 =0,~&~\forall~ i|m_i\neq 0.} 
}
These equations are equivalent to~\betag.

Consider the  {\it unperturbed} quiver
theory (with all $m_i=0$). This theory has $2n$ complex couplings
$(\tau_i,\lambda_i)$ and~\betacoup\ represent $n$ constraints on
them. Recalling the analysis of the $\CN=4$ theory in~\RGLMJS, we can
imagine solving these constraints to give the $\lambda_i$ as functions
of the $\tau_i$.  Thus we recover a moduli space,
$\CM^{(n)}_\tau$, whose 
global structure 
has been described by~\refs{\WittenSC,\DoreyQJ}. It is the moduli
space of an 
elliptic curve with modulus $\tau = \sum \tau_i$ and $n-1$ marked
points corresponding to the remaining independent couplings 
$\tau_i-\tau_{i+1}$. 

We now add $\mu$ mass perturbations. There are now $2n+\mu$
complex couplings $(\tau_i,\lambda_i,m_i)$, but~\betacoup\ still 
represent only $n$ constraints. We can again use the constraints to
eliminate the $\lambda_i$ in favor of the $(\tau_i,m_i)$.   We
recover a 
space of fixed points with the structure $\CM^{(n)}_\tau\times
\BP^{\mu-1}$, where the $\BP^{\mu-1}$ is parameterized by the ratios
$m_i/m_j$, as discussed above. We find superconformal fixed points at
each point of the  $\BP^{\mu-1}$. It is easy to see that the
$\BP^{\mu-1}\subset \BP^{n-1}$ is the vanishing locus of $n-\mu$ of
the $m_i$ in the theory with all $n$ masses turned on. 

In~\BergmanQI, a field theoretic computation of the central charges
was made when all $m_i\neq 0$. In general, the relations~\betacoup\ on
the vanishing loci of the $m_i$ do not lend a general result for the
central charge. However, the most obvious solution of~\betacoup, given
by~\GubserVD\ $\gamma_{\Phi_i}=1/2$ and
$\gamma_{A_i}=\gamma_{B_i}=-1/4$, leads to the result 
${c_{IR} \over c_{UV}} = {27 \over 32}$. We
conjecture that this is the central charge over the whole
$\BP^{n-1}$, {\it including the vanishing loci},  but we do not
have a field-theoretic proof of this.   

\newsec{The Massless Spectrum of IIB Strings on AdS$_5\times S^5/\BZ_n$} 

We now turn to the gravity dual to the quiver gauge theory on
D3-branes at $\BC^2/\BZ_n$.  Our ultimate goal is to identify the duals
of the flows described in the previous section.  To accomplish this
we first need to identify the duals of the operators in the perturbations
\Genopdef\ that drive the flow, and this is done by studying the linearized
perturbations about the UV fixed point theory.

As explained in~\KachruYS, the near horizon geometry of
D3-branes at $\BC^2/\BZ_n$ is AdS$_5\times S^5/\BZ_n$, where the 
$\BZ_n$ acts as a subgroup of $SU(2)_L$. Specifically, if the $S^5$ is
 embedded in $\BC^3$ given by $(z_1,z_2,z_3)$, then the
 $\BZ_n$ action is
\eqn\orbaction{(z_1,z_2,z_3)\longrightarrow  
(\alpha\, z_1, \alpha^{-1} \,z_2,z_3),}
where $\alpha=e^{2\pi i/n}$ is a root of unity.
We are interested in perturbations of the quiver gauge theory by
relevant operators. In the gravity dual, these operators correspond to
certain ``negative-mass\foot{ Despite their
$m^2<0$ Kaluza-Klein eigenvalues, these modes propagate on the
light-cone in AdS$_5$ and so one properly calls them massless.}''
modes on $AdS_5$.   
The spectrum of the orbifold theories was
given in~\refs{\OzHR,\GukovKK} (see also~\MorrisonCS\ for additional
discussion), and we now review those results 
starting with the simplest part: the untwisted sector.

\subsec{Untwisted Sector}

The untwisted sector will come from $\BZ_n$-invariant harmonics of the 
reduction of IIB 
supergravity on $S^5$~\OzHR.  Relevant operators correspond to modes with 
mass $m^2\leq 
0$. These can be obtained from the reduction of the $\CN=8$ graviton 
supermultiplet, as 
summarized in Table~3.1.
\medskip
\vbox{\ninepoint{
$$
\vbox{
\offinterlineskip
\def\tablerule{\noalign{\hrule}}
\halign{ 
\strut\vrule  # & \vrule \hfil #\hfil  & \vrule \hfil #  \hfil \vrule 
\tabskip=0pt\cr\tablerule \hskip5pt Field & $SU(4)$ irrep & $SU(2)_L\times 
SU(2)_R\times 
U(1)$ decomposition \cr \tablerule & ${\bf 20'}$ & $({\bf 1},{\bf 1})_0 
\oplus ({\bf 
1},{\bf 1})_4 \oplus ({\bf 1},{\bf 1})_{-4} \oplus ({\bf 2},{\bf 2})_2 
\oplus ({\bf 
2},{\bf 2})_{-2}\oplus ({\bf 3},{\bf 3})_0$ \cr \hskip5pt scalars & 
${\bf 10_c}$ & 
$2({\bf 2},{\bf 2})_0 \oplus ({\bf 3},{\bf 1})_2 \oplus ({\bf 3},{\bf 1})_{-2} 
\oplus({\bf 1},{\bf 3})_2\oplus ({\bf 1},{\bf 3})_{-2}$ \cr & ${\bf 1_c}$ & 
$2({\bf 
1},{\bf 1})_0$ \cr \tablerule & ${\bf 20}$ & $({\bf 2},{\bf 1})_{-1} 
\oplus ({\bf 
2},{\bf 1})_{3} \oplus ({\bf 1},{\bf 2})_{-3} \oplus({\bf 1},{\bf 2})_{1}  
\oplus  ({\bf 
3},{\bf 2})_{1} \oplus  ({\bf 2},{\bf 3})_{-1}$ \cr \hskip5pt fermions & 
${\bf 20^*}$ & 
$({\bf 2},{\bf 1})_{1} \oplus ({\bf 2},{\bf 1})_{-3} \oplus 
({\bf 1},{\bf 2})_3 
\oplus({\bf 1},{\bf 2})_{-1}  \oplus  ({\bf 3},{\bf 2})_{-1} \oplus 
 ({\bf 2},{\bf 
3})_1$ \cr & ${\bf 4}$ & $({\bf 2},{\bf 1})_{1} 
\oplus ({\bf 1},{\bf 2})_{-1}$\cr & 
${\bf 4^*}$ & $({\bf 2},{\bf 1})_{-1} \oplus ({\bf 1},{\bf 2})_{1}$\cr 
\tablerule 
\hskip5pt vectors & ${\bf 15}$ & $({\bf 1},{\bf 1})_0 
\oplus ({\bf 2},{\bf 2})_2 \oplus 
({\bf 2},{\bf 2})_{-2} \oplus ({\bf 3},{\bf 1})_0 
\oplus ({\bf 1},{\bf 3})_0$ \cr 
\tablerule \hskip5pt 2-forms & ${\bf 6_c}$ &  $2({\bf 1},{\bf 1})_2 
\oplus 2({\bf 
1},{\bf 1})_{-2}\oplus 2({\bf 2},{\bf 2})_0$\cr \tablerule \hskip5pt 
gravitini & ${\bf 
4}$ & $({\bf 2},{\bf 1})_1 \oplus ({\bf 1},{\bf 2})_{-1}$\cr & ${\bf 4^*}$ 
& $({\bf 
2},{\bf 1})_{-1} \oplus ({\bf 1},{\bf 2})_{1}$\cr \tablerule \hskip5pt 
graviton & ${\bf 
1}$ & $({\bf 1},{\bf 1})_0$ \cr \tablerule }}
$$
\noindent{\ninepoint\sl \baselineskip=8pt {\bf Table~3.1}:
The massless fields of IIB supergravity on 
AdS$_5\times S^5$ and the relevant branching rules.   }
}}
\medskip

The isometry group of $S^5$ is $SO(6) \approx SU(4)$. 
The $\BZ_n$-action of~\orbaction\ acts as the $SU(2)_L$ component of the
subgroup   $SU(2)_L\times SU(2)_R\times U(1) \subset SU(4)$. 
The relevant branching rules of 
$SU(4)\rightarrow SU(2)_L\times SU(2)_R\times U(1)$  are 
obtained from:
\eqn\basicbranch{ 
 {\bf 4} \rightarrow ({\bf 2},{\bf 1})_1 \oplus 
({\bf 1},{\bf 2})_{-1}  \,, \qquad 
 {\bf 4^*} \rightarrow ({\bf 2},{\bf 1})_{-1} \oplus 
({\bf 1},{\bf 2})_{1}\,. }
For the  ${\bf 6} = ({\bf 4} \otimes {\bf 4})_a$, one then has:
\eqn\vectbranch{
{\bf 6} \rightarrow ({\bf 1},{\bf 1})_2 \oplus 
({\bf 1},{\bf 1})_{-2}\oplus ({\bf 2},{\bf 2})_0\,.}
Continuing this, the ${\bf 20^*}$  appears in the product ${\bf 4^*}
\otimes{\bf 6} = 
{\bf 4}\oplus {\bf 20^*}$, and so one obtains  :
\eqn\twentystarbranch{ {\bf 20^*} \rightarrow ({\bf 2},{\bf 1})_{1} 
\oplus ({\bf 2},{\bf 
1})_{-3} \oplus ({\bf 1},{\bf 2})_3 \oplus({\bf 1},{\bf 2})_{-1}  
\oplus  ({\bf 3},{\bf 
2})_{-1} \oplus  ({\bf 2},{\bf 3})_1. }
It is then easy to read off the states from Table~3.1 which survive 
the 
$\BZ_n$ projection. The only subtlety is that the $\BZ_2$ projection, 
unlike the 
$\BZ_{n>2} $ projection, leaves invariant all the vector-like $SU(2)$ 
representations.   
For the gravitini we thus find 
$$
({\bf 2},{\bf 1})_1 
\hskip-1.2cm \vbox{\hrule height3.8pt width1.1cm depth-3.6pt}
\hskip.1cm \oplus
({\bf 2},{\bf 1})_{-1} 
\hskip-1.4cm \vbox{\hrule height3.8pt width1.1cm depth-3.6pt}
\hskip.3cm \oplus
({\bf 1},{\bf 2})_{-1}  \oplus 
({\bf 1},{\bf 2})_{1} 
\longrightarrow
{\bf 2}_{-1}  \oplus 
{\bf 2}_1,
$$
where the labels on the right-hand side are those
of $SU(2)_R\times U(1)$. For the vectors we find
$$
({\bf 1},{\bf 1})_0 
\oplus  ({\bf 2},{\bf 2})_2 
\hskip-1.2cm\vbox{\hrule height3.8pt width1.1cm depth-3.6pt}
\hskip.1cm\oplus ({\bf 2},{\bf 2})_{-2} 
\hskip-1.4cm\vbox{\hrule height3.8pt width1.1cm depth-3.6pt}
\hskip.3cm\oplus 
({\bf 3},{\bf 1})_0 \oplus ({\bf 1},{\bf 3})_0
\longrightarrow 
{\bf 1}_0  \oplus
(1+2\delta_{n2})({\bf 1}_0)\oplus {\bf 3}_0.
$$
We see that there are two states which are $\BZ_2$-invariant,
but not $\BZ_{n>2}$-invariant.  This leads to a pair of extra vector
multiplets for  $\Gamma = \BZ_2$ because $\BZ_2$ is, of course,
the center of $SU(2)_L$.

The set of fields which survive the $\BZ_n$ projection are displayed in 
Table~3.2.   For completeness we will also consider the $SU(2)_L$
projection used in \KPW.   The stronger $SU(2)_L$ projection of \KPW\ 
 further  truncates the states, since triplets of $SU(2)_L$ will give rise 
to a singlet 
of $\BZ_n$, leading to a single vector multiplet when $n>2$.   In
addition, all of the $SU(2)_L$ triplets are $\BZ_2$ invariant, so there
are three vector multiplets when $n=2$. Terms in brackets in Table~3.2
are $\BZ_{n>2}$ singlets that arise from  
$SU(2)_L$ triplets;  the additional states that are present for   
$\Gamma = \BZ_2$ are 
listed in the last column.  

\medskip
\vbox{\ninepoint{
$$
\vbox{
\offinterlineskip\tabskip=0pt
\def\tablerule{\noalign{\hrule}}
\halign{ 
\strut\vrule  # & \vrule \hfil #\hfil  & \vrule \hfil #  \hfil \vrule 
\cr\tablerule
\hskip5pt Field &  $SU(2)_R\times U(1)$ irrep & Additional states for 
$\Gamma = \BZ_{2}$ \cr 
\tablerule
& ${\bf 1}_0 \oplus {\bf 1}_4 \oplus
{\bf 1}_{-4} \bigl[\oplus {\bf 3}_0\bigr]$ 
& $\oplus 2({\bf 3}_0)$\cr
\hskip5pt scalars & ${\bf 3}_2\oplus {\bf 3}_{-2}
\bigl[\oplus {\bf 1}_2 \oplus {\bf 1}_{-2}\bigr]$
& $\oplus 2({\bf 1}_2 \oplus {\bf 1}_{-2})$ \cr
& $2 ({\bf 1}_0)$ &  \cr
\tablerule
\hskip5pt fermions & ${\bf 2}_3\oplus{\bf 2}_{-3} \oplus {\bf 2}_1 
\oplus  {\bf 2}_{-1} 
\bigl[ \oplus {\bf 2}_1 \oplus  {\bf 2}_{-1}\bigr]$ 
& $\oplus2({\bf 2}_1 \oplus  {\bf 2}_{-1})$\cr
& ${\bf 2}_{1}\oplus {\bf 2}_{-1}$ & \cr
\tablerule
\hskip5pt vectors & ${\bf 1}_0 \oplus {\bf 3}_0 
\bigl[\oplus{\bf 1}_0\bigr]$ 
& $\oplus2({\bf 1}_0)$  \cr
\tablerule
\hskip5pt 2-forms &  $2({\bf 1}_2 )\oplus 2({\bf 1}_{-2})$& \cr
\tablerule
\hskip5pt gravitini & ${\bf 2}_{-1} \oplus {\bf 2}_{1}$& \cr
\tablerule
\hskip5pt graviton & ${\bf 1}_0$ & \cr
\tablerule
} }
$$
\noindent{\ninepoint\sl \baselineskip=8pt {\bf Table~3.2}:
The massless fields in the untwisted sector of  IIB string
theory on $AdS_5\times S^5/\BZ_n$.  }
}}
\medskip
 
From \GRW\ we know that the gauged $\cN=4$ supersymmetry multiplets
have the content displayed in Table~3.3.

\medskip{\vbox{\ninepoint{$$\vbox{\offinterlineskip\tabskip=0pt
\halign{\strut\vrule #&\hfil   # 
&\hfil  # \hfil &  \vrule # 
&\hfil  # \hfil & \vrule #
&\hfil  # \hfil & \vrule # 
&\hfil  # \hfil & \vrule # \cr\noalign{\hrule}
&   &    &   & graviton  &   & tensor &   & vector  &  \cr
&   &    &   &~multiplet~ &   & ~multiplet~ &  & ~multiplet~   &  
\cr\noalign{\hrule}
&   & graviton  & & ${\bf 1}_0$  &   &   &   &   &  \cr\noalign{\hrule}
&  & gravitini  &  & ${\bf 2}_{-1} \oplus {\bf 2}_{1}$  &   &   &   &    &  
\cr\noalign{\hrule}
&   & 2-tensors  &   & $ {\bf 1}_2 \oplus  {\bf 1}_{-2}$  &   & ${\bf 1}_0$  
& &    &  \cr\noalign{\hrule}
&  & vectors  &   & ${\bf 1}_0 \oplus {\bf 3}_0$  &   &  &   & ${\bf 1}_0$  
&  \cr\noalign{\hrule}
&   & fermions  &  & ${\bf 2}_{-1} \oplus {\bf 2}_{1}$  &   & ${\bf 2}_{-1}
 \oplus {\bf 2}_{1}$ &   & ${\bf 2}_{-1} \oplus {\bf 2}_{1}$  &  \cr
\noalign{\hrule} &    & scalars  &   & ${\bf 1}_0$  &   & ${\bf 1}_{-2} \oplus {\bf 
3}_0  \oplus {\bf 1}_{2} $ &   &${\bf 1}_{-2} \oplus {\bf 3}_0  \oplus {\bf 1}_{2} $ 
&     \cr\noalign{\hrule}}\hrule}$$\vskip-7pt\noindent{\bf Table~3.3}:
Field content of   
the supermultiplets of the  $\cN=4$  $AdS_5$ superalgebra $SU(2,2|2)$ with the 
R-symmetry group $SU(2)_R\times U(1)_R \subset USp(4)$. The subscripts
denote the  
$U(1)_R$ charges and differ from the gauged $U(1)$ charges , as given
for example in  
Table 3.2. The gauged $U(1)$ is the diagonal subgroup of $U(1)_R$ and
another $SO(2)$  
symmetry  of the supergravity theory. }\vskip7pt}}

\noindent
From this we can read off the field content of the various
invariant  sectors of the $\cN=8$ theory:
 
\item{$\bullet$} $SU(2)_L$ invariant sector:   A graviton supermultiplet and
two charged tensor multiplets, of charges $\pm 2$ with respect to the gauged 
$U(1)$.   The scalar coset is  
${SO(5,2) \over SO(5)\times SO(2)} \times SO(1,1)$. 
\item{$\bullet$}  $U(1)_L$ {\it or} $\BZ_{n>2}$ invariant sector: 
A graviton 
supermultiplet, two 
charged tensor multiplets, of charges $\pm 2$, and a neutral vector 
multiplet.   The 
scalar coset is  ${SO(5,3) \over SO(5)\times SO(3)} \times SO(1,1)$. 
No fields are 
charged under the   vector field in the additional vector multiplet.
\item{$\bullet$}  $\BZ_2$ invariant sector: A graviton supermultiplet, 
two charged tensor multiplets, of charges $\pm 2$, and three  
vector multiplets.   
The scalar coset is  ${SO(5,5) \over SO(5)\times SO(5)} \times SO(1,1)$. 
The three 
vector multiplets actually gauge $SU(2)_L$, and this sector 
represents $\cN=4$ 
gauged supergravity with an $SU(2)_{L} \times SU(2)_{R} \times U(1)$ 
gauge group.

\subsec{The Twisted Sector}

The twisted sector states were computed in~\GukovKK\foot{A completely stringy 
computation of the spectrum was made in~\BerensteinHY.}. They are localized 
at the fixed 
circle of the $\BZ_n$ action~\orbaction\  (they propagate on AdS$_5\times 
S^1$), so they 
can be computed via KK-reduction of the IIB string theory on $S^1 \times 
\CM$. Here 
$\CM$ is a compact Einstein manifold which looks locally like 
$\BC^2/\BZ_n$, but 
inherits curvature and 5-form flux (such that $\Lambda=4$ in units of
the AdS$_5$ radius) from its 
relation to 
$S^5/\BZ_n$.   

The twisted sector states of IIB string theory on $\BC^2/\BZ_n$
consist of $(n-1)$ six-dimensional $(0,2)$ tensor multiplets. The field 
content of
the tensor multiplets is a tensor in the singlet of the $USp(4)$ $(0,2)$
R-symmetry, a set of chiral fermions in the ${\bf 4}$, and real
scalars in the ${\bf 5}$. The conformal field theory on this
background enjoys a $\BZ_n$ quantum symmetry~\VafaIH, which acts by
phases on the twisted sector states. In the theory on branes at the
singularity, the quantum symmetry acts by clock shifts on the quiver
diagram~\refs{\WittenKZ,\BerensteinDW}. 

The spectrum on  $\CM$ will similarly consist of tensor multiplets,
but there will be   
corrections to the masses of these states.  Reducing on the $S^1$ to
AdS$_5$ we thus get  
tensor multiplets, which, as we will discuss in section~4, are
not equivalent to vector  
multiplets in AdS$_5$. Moreover, due to the mass corrections,  it turns
out that the massless states in AdS$_5$ are tensors corresponding to
non-trivial KK  harmonics on the $S^1$.   This means that the massless
tensor multiplets in  
five dimensions come in charge-conjugate pairs under the $U(1)$
associated to the Hopf  
fiber.  In terms of physical 
degrees of freedom, the scalars in the ${\bf 3}_2$ correspond to the
hyper-K\"ahler  
deformations of $\BC^2/\BZ_n$. The two singlets correspond to the
periods of the NS-NS  
and RR B-fields. We have summarized the twisted sector fields in Table~3.4.

One should note that these twisted sector tensor multiplets have
precisely the same  
charge assignments as the pair of tensor multiplets that come from the
untwisted sector.  
As we will see, this is an essential consequence of the $\BZ_n$ cyclic
quantum  symmetry of the quiver theory.

\medskip
\vbox{\ninepoint{
$$
\vbox{\offinterlineskip
\def\tablerule{\noalign{\hrule}}
\halign{ 
\strut\vrule  # & \vrule \hfil #\hfil  \vrule
\cr 
\tablerule
\hskip5pt Field &  $SU(2)_R\times U(1)$ irrep \cr
\tablerule
\hskip5pt scalars & ${\bf 1}_4 \oplus {\bf 3}_2 \oplus {\bf 1}_{0}$ \cr
& ${\bf 1}_{-4} \oplus {\bf 3}_{-2} \oplus{\bf 1}_0$ \cr
\tablerule
\hskip5pt fermions & ${\bf 2}_3 \oplus {\bf 2}_{1}
\oplus{\bf 2}_{-1} \oplus {\bf 2}_{-3}$\cr
\tablerule
\hskip5pt 2-forms &  ${\bf 1}_2\oplus {\bf 1}_{-2}$ \cr
\tablerule
}}
$$
\noindent{\ninepoint\sl \baselineskip=8pt {\bf Table~3.4}:
The massless fields in the twisted sector of  IIB string
theory on $AdS_5\times S^5/\BZ_n$ consist of $n-1$ copies of these
fields.  } 
}}
\medskip

\subsec{Dual Operators in the Quiver Gauge Theory} 

To identify the states above with operators in the field theory, it is
important to recall the discussion of the $\cN=2$ superconformal
R-charge in section 2. We denote the scalar components of the
bifundamental chiral multiplets as $(a_i,b_i)$ and those in the adjoint
multiplets by $\phi_i$. The pairs $(a_i,\bar{b}_i)$ and $(b_i,\bar{a}_i)$
form doublets under $SU(2)_R$, and are neutral under the
$U(1)$;   the fields, $\phi_i$, are $SU(2)_R$ singlets with
$U(1)$-charge $+2$. 

It is thus elementary to identify the duals of $\phi_i^2$ 
and $(\bar{\phi}_i)^2$ with the   ${\bf 1}_{+4}$
and ${\bf 1}_{-4}$ of the tensor multiplets.\foot{Throughout this
discussion we have suppressed explicit traces on
products of operators: One
should remember that all the operator expressions contain
gauge invariant traces.}  In the untwisted sector these tensor
multiplet scalars are dual to
the surviving residue of the corresponding operators
in the original  $\cN=4$ theory.  That is, they are
dual to  $\sum_i^n \phi_i^2$  and $\sum_i^n (\bar{\phi}_i)^2$,
which are, of course, singlets under the quantum $\BZ_n$ 
symmetry.  The scalars in the twisted sector are, of course,
dual to operators that are charged under the quantum $\BZ_n$
symmetry, and which are orthogonal to diagonal sums
like  $\sum_i^n \phi_i^2$.    It is therefore convenient to
use a cyclic basis $\phi_i^2-\phi_{i-1}^2$.
Similarly, one should recall that  the ${\bf 1}_{0}$ scalars in the 
tensor multiplet of the twisted sector
come from the periods of the NS-NS and RR $B$-fields over 
the blown-down $\BP^1$ cycles of the ALE space.  They are dual
to {\it differences} between the complexified gauge couplings of the
quiver gauge theory~\refs{\KachruYS,\LawrenceJA}.  The
sum over all the gauge couplings is dual to the original
IIB dilaton, which is now in the tensor multiplet of the
untwisted sector.

The $SU(2)_R$ triplets are easily identified by
using the superconformal algebra.  They are fermion
bilinear operators for the two fermions in the $\cN=2$ 
vector multiplet:  $\cO_i^{ab}  \equiv \chi^a_i \chi^b_i$.  
As above, the scalars in the untwisted sectors are dual
to the sums, $\sum_i^n \cO_i^{ab}$, while the scalars in the
twisted sectors are dual to the differences 
$\cO_i^{ab} -\cO_{i-1}^{ab}$.

We have thus identified the duals of operator bilinears
of the $\cN=2$ vector multiplets on the brane with charged
tensor multiplets in supergravity.  
One should note that having identified the dual of any
scalar in the tensor multiplet (such as the gauge coupling)
the entire holographic assignment follows from supersymmetry.
One can indeed check that this is consistent with the foregoing
identifications.  One can also arrive at the same result far more
generally: The gauge couplings must be ${\bf 1}_{0}$'s of the 
R-symmetry, and from Table 3.3 we see that this is only possible if we 
start from vector or tensor multiplets of charges $\pm 2$.
It was shown in \DallAgataVB\ that in a gauged
$\cN=4$ supergravity theory {\it only tensor multiplets}
can be charged under a  $U(1)$ gauge symmetry.
Hence we see that the bilinear operators in the vector multiplets
on the brane must be dual to scalars in tensor multiplets
of charge $\pm 2$.

Finally, we come to the rest of the untwisted sector of
the bulk theory, which contains the graviton supermultiplet and either one or 
three vector multiplets. The duality is straightforward to establish: One
simply projects $\cN=4$ Yang-Mills dual operators
onto the fields of the quiver theory.  The dilaton of the
supergravity multiplet belongs to the $SO(1,1)$ factor of the scalar manifold
\scalarM\ and comes
from the ${\bf 20'}$ of $SU(4)$. Its dual is easily identified
 \refs{\KPW,\FGPWa} as
\eqn\dildual{
\sum_i^n  \left[|\phi_i|^2 ~ - ~\coeff{1}{2}\, 
\left( |a_i|^2 + | b_i|^2 \right) \right]\,.}
The relative normalization reflects the traceless condition on
the original ${\bf 20'}$ of $SU(4)$, ensuring that~\dildual\ is chiral.

There is also a  $\cN=4$  vector multiplet common to all $\BZ_n$
projections.  We  
identify the ${\bf 3}_0$ state with the triplet components of the
product of the two  
$SU(2)_R$ doublets we discussed above. These are the operators
$\sum_i^n a_i b_i$,  
$\sum_i^n \bar{a}_i \bar{b}_i$, and the $J_3$-component $\sum_i^n
 ( |a_i|^2 - | b_i|^2 )$.   These are the same expressions
which appear in the D and F-flatness conditions and so they are frozen to
constant values in the $\CN=2$ vacuum.

For $\Gamma = \BZ_2$ we obtain two more  vector multiplets, giving
three vector  
multiplets that in fact gauge an additional $SU(2)$.  (In fact it is
$SU(2)_L$.)  The  
new vector multiplets are dual to the multiplets built on the
$\BZ_2$-invariant chiral  
primaries $a_1a_2$  and $b_1b_2$ and their $SU(2)_R$ images. More
generally, to form  
gauge invariant operators like these one must take products over all
the nodes of the  
quiver: $\prod_i^n a_i$ and  $\prod_i^n b_i $, and these have
dimension $n$.   They are  
thus only relevant supersymmetric perturbations for $n=2$. These
operators could be  
useful to probe the relative anomalous dimensions acquired by the
$A_i$ and $B_i$  
multiplets.

\newsec{ $\CN=4$ Gauged Supergravity Theories}

In the 1980's gauged supergravity theories were extensively
studied, but primarily in four dimensions.  The five-dimensional
theories were also studied, but less thoroughly.   The maximal
gauged, $\cN=8$ theory was constructed \refs{\GRW,\PerniciJU}, 
and 
some $\cN=4$ theories coupled to matter were investigated
(see, for example, \refs{\AwadaEP,\RomansPS}).  However, it
was only rather recently that the most general gauged
$\cN=4$ supergravity with matter coupling was constructed
 \DallAgataVB.  As we have discussed, the most general massless
matter multiplets that are consistent with  $\cN=4$ supersymmetry
are vector and tensor multiplets, and these are not equivalent in
the AdS backgrounds of gauged supergravity theories.

We  are interested in these theories here because we wish
to find simple descriptions of RG flows in quiver gauge theories.
This approach has already proven very successful in $\cN=4$
Yang-Mills using $\cN=8$ supergravity, and the successes rested
heavily on the five-dimensional theory being a consistent truncation
of IIB supergravity.  Consistent truncation in $\cN=4$ supergravity
is largely {\it terra incognita}, but we will start the discussion with
results that are well established.  To go beyond these, it is worth
remembering 
that one of the reasons, and perhaps the primary reason, why it
works is the complete rigidity of the structure.  The complete
$\cN=8$ theory is determined by the perturbative spectrum and
the choice of gauge group.  In terms of the brane theory this
means that the large $N$ operator product structure within the
energy-momentum tensor supermultiplet is rigidly determined
by supersymmetry and $R$-symmetry.   It was shown in  \DallAgataVB\
that the entire structure of the $\cN=4$ supergravity theory is
fixed once one has decided on the gauge group and how the
tensor multiplets are charged under this gauge group.
This rigidity of structure should therefore be reflected in the
large $N$ operator product structure of the $\cN=2$ theory on 
the brane, and thus determine the RG flows whether one
uses the five-dimensional, or the ten-dimensional descriptions.

The bottom line is, as we described in the introduction,  we find a 
family of flow solutions  in the five-dimensional theory that exactly 
match onto the known ten-dimensional fixed-point solutions.

\subsec{Terra Cognita: Truncations of the $\cN=8$ theory} 

It is an often used fact that it is always consistent to truncate any
field theory to the sector of singlets  under any symmetry,
discrete or continuous.   One may thus generate several 
gauged $\cN=4$ theories by applying this technique to
the $\cN=8$ theory, and the simplest way to accomplish this 
is to look for singlets under some subgroup, $\Gamma 
\subset SU(2)_L$ in the embedding $SU(2)_L  \times 
SU(2)_R \times U(1) \subset SO(6)$.  The whole point is that
$\Gamma$ singlets contain four of the eight  original supersymmetries.

The obvious choices for $\Gamma$ are 
$SU(2)_L$, $U(1)_L\subset SU(2)_{L}$ and $\BZ_n$, and as we have
discussed, they lead to $\cN=4$ supergravity coupled to
two charged tensor multiplets, and either zero, one or three
vector multiplets.   The resulting theories are perfectly consistent
closed subsectors of the $\cN=8$ supergravity theory, and any
result derived therein can be, at least in principle, unambiguously
lifted to ten dimensions.  

In terms of the quiver theories on the brane, these $\cN=4$
theories represent the untwisted sector of the quiver theory, and
flows within these $\cN=4$ theories may be interpreted
as  flows lying entirely within the untwisted sector of the
corresponding quiver.  That this is consistent may also be seen within
the field theory as a consequence of the $\BZ_n$ quantum symmetry.
It follows that the $\cN=1$ supersymmetric flows of
\FGPWa\ must be a part of any quiver theory, and indeed
this observation was made in \FGPWa.

We now wish to go beyond these results, and add in precisely $(n-1)$
pairs of tensor  
multiplets with charges $\pm 2$ under the $U(1)$ factor of the
$SU(2)_R \times U(1)$   
gauge group. 

\subsec{$\cN=4$ gauged supergravity theories: Some general
facts} 

The starting point for constructing a gauged supergravity theory is
usually to start  
with the ungauged theory in Minkowski space.  In this setting the
vector and tensor  
fields are equivalent, and so we may start with the theory described
in \AwadaEP,  
consisting of the $\cN=4$ Poincar\'{e} 
supergravity multiplet coupled to an arbitrary number, $p$, of 
vector multiplets. 

The five-dimensional $\cN=4$  Poincar\'{e} supergravity
multiplet  consists of a  
graviton, four gravitini,  six vector fields, four ``spin-$\half$''
fields and a single scalar (the dilaton). There is an $SO(5) \cong
USp(4)$, R-symmetry with the spinors  
transforming as   a $\bf 4$ and the vectors transforming as a $\bf 5 +
1$ under it.  In the ungauged Poincar\'{e} supergravity theory
the massless  ``matter multiplets''  can only be vector 
multiplets\foot{Additional  
gravitino multiplets would require more supersymmetry.}, and such a
multiplet contains  one vector field, four spin-$\half$ fields and
five real scalars. 

The addition of the $p$ vector multiplets to supergravity thus results in
the bosonic field content:
\eqn\fields{
\{ e_{\mu}^{m}, a_{\mu}, A_\mu^{\tilde{I}}, \sigma, \phi^{x} \},}
where $e_{\mu}^{m}$ denotes the f\"{u}nfbein, $a_{\mu}$ is the
$SO(5)$ singlet vector field of the supergravity multiplet, the
$A_{\mu}^{\tilde{I}}$ $(\tilde{I}=1,\ldots,(5+p))$ comprise the
$SO(5)$ five-plet of vector fields from the supergravity multiplet as
well as the $p$ vector fields from the $p$ vector multiplets,
$\sigma$ is the supergravity scalar (the ``dilaton"), 
and $\phi^{x}$
$(x=1,\ldots,5p)$ collectively denotes the scalar fields of the
vector multiplets.

The scalar fields $(\phi^{x},\sigma)$ parameterize the scalar
manifold
\eqn\scalarM{
\CS={SO(5,p) \over SO(5)\times SO(p)}\times SO(1,1) ,}
where the coset part and the $SO(1,1)$ factor are due to $\phi^{x}$ and
$\sigma$, respectively.

The isometry group $G=SO(5,p)\times SO(1,1)$ of  $\CS$
extends to a rigid symmetry of the entire ungauged supergravity
Lagrangian. Under this symmetry group, the vector fields
$A_{\mu}^{\tilde{I}}$ transform in the ${\bf (5+p)}$ of
$SO(5,p)$ and have $SO(1,1)$ charge $-1$, while $a_\mu$ is
$SO(5,p)$ inert and has $SO(1,1)$ charge $+2$.

A gauged version of this theory is obtained by ``gauging''
appropriate subgroups $K\subset G$.  For the gauged $\cN=4$
supergravity   with $\cN=4$ supersymmetric AdS ground states, 
one must consider the corresponding supermultiplets of the 
$\cN=4$ anti-de Sitter $AdS_5$  superalgebra
$SU(2,2|2)$ with an R-symmetry group $SU(2)_R\times 
U(1)_R \subset USp(4)$. The $\cN=4$,
$AdS_5$ graviton supermultiplet consists of the graviton, 
four gravitini, $(3+1)$ vector fields, two   tensor fields and 
one scalar (the dilaton). Technically, the appearance of the two  
tensor fields can be traced back to the gauging of the $U(1)_R$ factor
in the   supergravity theory \RomansPS, which is a generic 
phenomenon in five-dimensions
\refs{\GRW,\PerniciJU,\GZ,\DallAgataVB} that also carries over to the
matter multiplets:   vector and tensor multiplets are no longer 
equivalent in gauged five-dimensional  supergravity. Indeed,  in the 
gauged theory, tensor fields  satisfy a first order system of
self-duality equations and  
these equations require that there be an even number of tensor
multiplets.  Thus  a generic gauged theory will contain an {\it even}
number, $2n$, of tensor  multiplets 
and   {\it an arbitrary number}, $q$, of vector multiplets.   The fermion and
scalar content of   
vector and tensor multiplets is the same, and the generic gauged
supergravity theory has  
a scalar coset of the form \scalarM\ with $p= 2n + q$.

Supersymmetry imposes severe
constraints on the possible gauge groups $K$. For our purposes,
the most important results of the general analysis of \DallAgataVB\ are:

\item{$\bullet$}
The $SO(1,1)$ factor in $G$ {\it cannot}  
be gauged, {\it i.e.}, any gauge group $K$ has to  be a subgroup of
$SO(5,p)$.

\item{$\bullet$}
When the gauge group, $K$, is Abelian, it has to be
{\it one}-dimensional with the corresponding gauge field given by
the $SO(5,p)$ singlet $a_{\mu}$. Furthermore, in the
gauged theory there can be  {\it no} vector fields charged
under this Abelian gauge group:  If one wants charged matter
fields then they must be tensor multiplets.  In the traditional
supergravity description one takes $K$ to be a subgroup
of $SO(5,p)$, and any vector field that is charged under $K$
must be dualized to a tensor field prior to gauging.
At the linearized level, the field equations of
such self-dual tensor fields are of the form \TPvN
$$
dB=im (\ast B),
$$
where $\ast$ denotes the Hodge dual, and $m$ is a mass parameter
proportional to the coupling constant for $K$.  Note that  
five-dimensional self-duality requires the
factor  $i$ on the right hand side. This implies that the tensor
fields have  to be complex and hence of even number when 
split in real and imaginary parts.

Note that the converse  is also true:  In five-dimensional $\CN=4$
gauged supergravity any self-dual tensor field has to be charged
with respect to a one-dimensional Abelian group $K\subset SO(5,p)$
\DallAgataVB. 

\item{$\bullet$}
When the gauge group $K$ is semi-simple, no such self-dual tensor
fields can exist, and the ${\bf (5+p)}$ of $SO(5,p)$ has to
decompose with respect to $K\subset SO(5,p)$ as
\eqn\Kdecomposition{
{\bf (5+p)} \longrightarrow {\rm adjoint}(K)\oplus
{\rm possible}~ K {\rm  -singlets}. }
In particular, there must not be any non-singlets of $K$ in
addition to the adjoint. Otherwise, the gauging would be
inconsistent with supersymmetry \DallAgataVB.

When the gauge group $K\subset SO(5,p)$ is a direct product of a
semi-simple and an Abelian group (as will be the case in all of the
examples considered here), a combination of the previous two items 
applies: the
Abelian factor has to be one-dimensional, its gauge field is
$a_{\mu}$, and the vector fields that would transform
non-trivially under this Abelian factor have to be converted to
self-dual tensor fields. Conversely, all self-dual tensor fields
have to be charged with respect to the Abelian factor, and they
must be neutral with respect to the semi-simple part of the gauge
group $K$. With respect to this semi-simple part, the
${\bf (5+p)}$ of $SO(5,p)$ again has to decompose as
in~\Kdecomposition.

\item{$\bullet$}
After $K$ has been gauged, the former global symmetry group $G=SO(5,p)
\times SO(1,1)$
is broken to $[SO(1,1)\times C]_{\rm rigid}\times
K_{\rm gauged}$, where $C$ denotes the commutant 
of $K$ in $SO(5,p)$ (modulo the Abelian
subgroup of $K$). When the gauge
group $K$ is the direct product of a semi-simple and an Abelian
factor, the global $SO(1,1)$ symmetry can be used to rescale the
ratio of the two coupling constants, so that only their
relative signs are physically meaningful \RomansPS.

\subsec{The scalar potential}

As we commented earlier, the structure of
the $\CN=4$ supergravity theories is almost completely rigid. 
Indeed, specifying how an allowed gauge group $K$ is embedded in
$SO(5,p)$   fixes the complete supergravity Lagrangian including the scalar 
potential.   The scalar structure, and most particularly the scalar
potential is of central importance to the study of the holographic flows,  
and so we now focus upon this sector in more detail.

We are interested in the case where the gauge group $K$ is a
direct product of a semisimple and an Abelian group. Denoting by
$g_{S}$ and $g_{A}$ the gauge couplings of, respectively,  the
semisimple and the Abelian factor, the general form of the scalar
potential reads \DallAgataVB
\eqn\potential{
\cV={1 \over 2}\,\Big[\; g_{A}^2\, V_{ij}^aV^{aij} -36\,
g_{A} g_{S} \, U_{ij}S^{ij} +g_{S}^2\,\Big(
 T_{ij}^aT^{aij}-9\, S_{ij}S^{ij}\Big)\,\Big]~.}

Here, the tensors $V_{ij}^{a}$, $U_{ij} $,  $T_{ij}^{a}$ and
$S_{ij} $
 are appropriately contracted products of two or three
 coset representatives of $SO(5,p)/[SO(5)\times
SO(p)]$. Denoting such a coset representative by
$L_{\tilde{I}}^{A}=(L_{\tilde{I}}^{ij},L_{\tilde{I}}^{a})$  (with
inverse $L_{A}^{\tilde{I}}$), where
$\tilde{I}$ is an $SO(5,p)$ index, $a$ is an $SO(p)$ index, and
$ij$ denotes the ${\bf 5}$ of $SO(5)$ written in terms of
$USp(4)$ indices $i,j, \ldots=1,\ldots, 4$ ({\it i.e.},
$L_{\tilde{I}}^{ij}$ is antisymmetric and symplectic traceless in
$i$ and $j$), these quantities are
\eqn\STUVtens{\eqalign{
U_{ij}=& {\sqrt2 \over 6}
e^{{2\sigma \over \sqrt{3}}}{\Lambda^{N}}_{M} L_{Nik}
L^{Mk}_{\,\,\,\,\,\,\,\,\,j}\cr
V_{ij}^{a}=&{1 \over \sqrt{2}}e^{{2\sigma \over \sqrt{3}}}
{\Lambda^{N}}_{M} L_{Nij}
L^{Ma}\cr
S_{ij} =& -{2 \over 9} e^{-{\sigma \over \sqrt{3}}} L^{J}_{ik}
{f_{JI}}^K L_{K}^{kl}
L^{I}_{lj},\cr
T_{ij}^a =& - e^{-{\sigma \over \sqrt{3}}} 
L^{Ja}L^K{}_i^{\ k}{f_{JK}}^I L_{Ikj}~.
}}
Here, ${\Lambda^{N}}_{M}$ defines the action of the  Abelian factor of $K$
upon the tensor fields, and ${f_{IK}}^{J}$ are the structure constants of the
semi-simple part of the gauge group (that is, we have split the index 
$\tilde{I}$ into $M$ and $I$, corresponding to the splitting of the
vector fields 
$A_{\mu}^{\tilde{I}}$ of the ungauged theory into tensor $B^M_{\mu\nu}$ and
vector fields $A^I_{\mu}$ \DallAgataVB\ in the gauged
version). 

As mentioned earlier, the global $SO(1,1)$ symmetry, which acts as a
constant shift on $\sigma$,   can be used to
rescale the coupling constants $g_{A}$ and $g_{S}$ such that only
their relative sign matters.

Finally, by examining the contractions in the potential \potential\
and in the definitions of the  tensors in \STUVtens, one sees that the 
potential is invariant under two symmetry operations:
\item{(i)} The left multiplication of the matrix $L$ by the
elements of $SO(5,p)$ that commute with
${\Lambda^{N}}_{M}$ and that leave ${f_{IK}}^{J}$ invariant.
This is precisely the group $C_{rigid}\times K_{gauged}$ mentioned 
earlier.
\item{(ii)} Right multiplication by any matrix in $SO(5) \times
SO(p)$,  which  is the  composite local symmetry of the theory.

One can use the second of these symmetries to put $L$ in
the ``symmetric gauge,'' in which $L$ is the exponential
of a symmetric matrix with $5p$ purely non-compact 
generators:
\eqn\Lsymmetric{
L = \exp \left(  \matrix{  {\bf 0}_{5\times 5} & X \cr
X^{T}& {\bf 0}_{p\times p} \cr}
\right), }
Having gone to the symmetric gauge one
still has the symmetry (i), but to preserve the symmetric gauge
one may have to combine it with a compensating transformation
using symmetry (ii).

\subsec{The relevant gauged supergravity theory }

In holographic duality the R-symmetry on the brane becomes
the gauge symmetry   of the supergravity, and so we need to
gauge $SU(2)_R \times U(1)$.  The perturbative compactification
analysis of section 3 gives us the spectrum and charges of the additional
matter multiplets: we have the $\cN=4$ supergravity multiplet coupled
to $n$ pairs of tensor multiplets of charge $\pm 2$, and we have
either one ($n >2$) or three  ($n =2$) additional vector multiplets
which gauge an additional $SU(2)_{L}$. 

We will focus initially on the theory with one additional
vector multiplet, corresponding to the $\BZ_{n>2}$ orbifolds.  
The scalar fields parameterize the coset space\foot{This should be
compared with the scalar coset for the six-dimensional effective
theory of  the IIB string on an $A_{n-1}$ space,
which is $SO(5,n+1)/[SO(5)\times SO(n+1)]$. }
\eqn\gencoset{  SO(1,1)\times {SO(5,2n+1) \over SO(5)\times
SO(2n+1)}.}
The $SU(2)_R$ factor is embedded into $SO(5)\subset SO(5,2n+1)$ as 
$ SO(3) \subset SO(3)\times SO(2) \subset SO(5)$, with 
${f_{IK}}^{J} = \epsilon_{IJK}$ in the first 
three indices of the $SO(5)$ vector representation.   The  embedding
of the $U(1)$ factor of $SU(2)_R 
\times U(1)$ is a little less obvious, but is the natural extension of
that found in  
\KPW. For the untwisted sector the coset has the numerator $SO(5,3)$,
and this contains  
$SO(5,2)$ in an obvious manner.  This contains 
$SO(3) \times SO(2,2) = SO(3) \times SU(1,1)  \times SU(1,1)$. One
$SU(1,1)$ represents the IIB dilaton, while the other  
$SU(1,1)$ represents geometric scalars coming from the $S^5$
metric. The $U(1)$ subgroup  
of this second $SU(1,1)$ is a geometric symmetry on the $S^5$ that
becomes a gauge  
symmetry in five dimensions.  This $U(1)$ sits inside $SO(2,2)$ as the
diagonal $SO(2)$  
in $SO(2) \times SO(2) \subset SO(2,2)$. Adding the twisted sector
tensor multiplets  
extends this to be the diagonal $SO(2)$ in 
$(SO(2))^{n+1} \subset SO(2,2n)$, and 
$SO(3)   \times SO(2,2n) \subset SO(5,2n) \subset SO(5,2n+1)$. Thus we
have: 
\eqn\abigmatrix{
{\Lambda^{N}}_{M}= \left(  \matrix{ {\bf 0}_{3 \times 3}  & 
0&0& 0& \dots& \dots & 0&0 \cr
 0 & \varepsilon&0&0& \dots& \dots&0&0\cr
0 & 0&\varepsilon&0& \dots&\dots&0&0\cr
0 & 0&0&\ddots& \ddots&\dots&0&0\cr
\vdots & \vdots&\vdots& \ddots&\ddots&\ddots&\vdots&\vdots\cr
0 & 0&0&\dots & \ddots&\ddots&0&0\cr
0 & 0&0& \dots&\dots&0&\varepsilon&0\cr
0 & 0&0& \dots&\dots&0&0&0\cr}
\right) \,,  \qquad \varepsilon \equiv 
\left(\matrix{0 &1 \cr -1 & 0}  \right) \,,
}
and where $\varepsilon$ is repeated $(n+1)$ times.

It is important to examine the symmetry of the corresponding
potential~\potential.
The symmetry from left-multiplication is $SU(2) \times
U(1,n)$.  The $SU(2)$  is the invariance of  ${f_{IK}}^{J}$,
while $U(1,n)$ is the subgroup of $SO(2,2n)$ that
commutes with ${\Lambda^{N}}_{M}$.  The $U(1)$ center
 of $U(1,n)$ is, of course, generated by $\Lambda$ itself,
and is part of the gauge symmetry.  If one passes to
symmetric gauge, then the potential in this gauge has 
the symmetries:
\item{(i)}  Conjugation  by $SU(2) \times U(1) \times U(n)$, which
is the compact subgroup of the symmetry group identified above.
\item{(ii)}  Left multiplication by $SU(1,n)$.  If one wishes
to preserve symmetric gauge~\Lsymmetric\ then this 
must be combined with a compensating right 
 multiplication by $U(1) \times U(n)$.

These symmetries are a very natural generalization of
the symmetries of the potential in the $\cN=8$ theory,
and the truncation considered in \KPW.  In particular
that truncation had an $SU(2) \times U(1) \times U(1)$, 
along with a non-compact $SU(1,1)$ symmetry that
corresponds to the IIB dilaton/axion coset.

We finish by noting that the model with three extra vector multiplets 
in the case $\Gamma=\BZ_{2}$ is a simple 
generalization of the foregoing. First  the supergravity gauge group
is now $SU(2)_L   \times SU(2)_R \times U(1)$, and so the structure
constants ${f_{IK}}^{J}$ can be taken  
to be $\epsilon_{IJK}$ in both the first three indices and the last
three indices of the  
$SO(5, 7)$ vector representation.  With this choice, the matrix
$\Lambda$ is as in  
\abigmatrix, but extended by two extra rows and columns of zeroes.
The potential still  
has the non-compact $SU(1,2)$ symmetry, but has the compact
conjugation symmetry of  
$SU(2) \times SU(2) \times  U(1) \times U(2)$.
 
\subsec{The five-dimensional scalars and the IIB fields}

Thus far we have identified holographic duals of
operators on the brane in terms of the supergravity scalars  
in $\cN=4$ multiplets.  Here we wish to make that identification
a little more explicit in terms of the five-dimensional 
scalar manifold.   It is well known that at the non-linear level the 
correspondence of fields can be quite subtle, but there is
also a minor issue at linearized level involving choices of basis.

We will focus on two particular components of the five-dimensional tensor
multiplets, namely the  ${\bf 1}_0$ and the $U(1)\subset SU(2)_R$
singlet part   
of the ${\bf 3}_{\pm 2}$ scalars. We will denote these complex degrees of
freedom as $\phi_i$ and $\chi_i$,  respectively.    In the
supergravity theory 
it is natural to use an orthonormal basis in terms of the Cartan-Killing 
form on $SO(5,2n+q)$, and so take a linearized kinetic
term of the form: 
\eqn\normkin{ \sum_{j=1}^n  \, (\partial_\mu \phi_j\,)^2 
 ~+~ \sum_{j=1}^n  \, (\partial_\mu \chi_j\,)^2  \,.}

In the linearized IIB theory the counterparts of the
$\phi_j$ and $\chi_j$,  which we will call $\hat{\phi}_j$ and 
$\hat{\chi}_i$, have very diverse origins in terms of untwisted 
and twisted sectors.  Consider the $\hat{\phi}_j$:
One of them, say $\hat{\phi}_1$, comes from the untwisted
sector and is the complexified axion-dilaton, which
corresponds to the {\it sum} of the complexified gauge couplings,
$\sum_i \tau_i$, in
the quiver theory.  In the twisted sector, the $\hat{\phi}_i$ are
proportional to the periods $B_{NS} + i B_{RR}$  over the homology
2-spheres of the ALE 
space.  They correspond to differences, $\tau_i - \tau_{i+1}$, between
the complexified gauge 
couplings of the quiver theory. The intersection form
on the 2-cycles is given by the
Cartan matrix, $C_{ij}$, of the $A$--$D$--$E$ group and so
$C_{ij}$ appears as a mixing matrix in
the effective action for the $\hat{\phi}_i$ . So at the linearized level, 
there is a natural field re-definition between $\phi_i$ and 
$\hat{\phi}_i$  to get a canonical kinetic term for the fields $\phi_i$:
\eqn\normkin{ \sum_{i,j=1}^n\delta^{ij} \, \partial_\mu \phi_i\,
\partial^\mu \phi_j 
= (\partial_\mu \hat \phi_1\,)^2 ~+~ \sum_{i,j=2}^n {C}^{ij} \, 
\partial_\mu \hat{\phi}_i\,  \partial^\mu
\hat{\phi}_j \,.} 

In other words, the twisted sector fields $\hat{\phi}_i$ are in
one--to--one correspondence with the simple roots $\alpha_i$ of the 
$A$--$D$--$E$ group. We choose a basis for the fields $\phi_i$ which
behaves like the fundamental weights $e_i$, which satisfy 
$e_i\cdot e_j =\delta_{ij}$, in order to obtain canonical kinetic
terms~\normkin. The twisted sector linearized fields $\hat{\phi}_i$
are proportional to differences of the $\phi_i$ through the expression
$\alpha_i = e_i - e_{i+1}$ for the simple roots in terms of the
weights. We further make the choice that the untwisted sector field
$\hat{\phi}_1= c \sum_{i=1}^n \phi_i$ for some normalization $c$. With
this choice of basis, the $\phi_i$ satisfy the right properties to be
dual to the coupling constants $\tau_i$ at each node of the quiver.
Similarly, the $\chi_i$ are dual to the mass perturbations $m_i$
in~\Genopdef. 

It is also useful to describe how these scalars sit in the
coset~\gencoset. One  
must examine $SO(2,2)$ subgroups of $SO(5,2n)$ to determine the $U(1)$
R-charges of the  
various generators~\KPW. One finds that the general representative
with all fields  
except $\phi_i$ and $\chi_i$ turned off takes the form 
\eqn\subsetrep{ L = \exp \left(  \matrix{  {\bf 0}_{5\times 5} &
v^{(1)}_{5\times 2} &  
\cdots  & v^{(n)}_{5\times 2} & {\bf 0}_{5\times 1}\cr
(v^{(1)}_{5\times 2})^T & &  
& &  \cr \vdots & & {\bf 0}_{2n+1\times 2n+1} & \cr (v^{(n)}_{5\times
2})^T & & & &   
\cr {\bf 0}_{1\times 5} & & & &  \cr} \right), }
 where the $5\times 2$ matrices have the form
\eqn\submatrix{
v^{(i)}_{5\times 2} = \left(
\matrix{ {\rm Re}~\chi_i & {\rm Im}~\chi_i \cr
- {\rm Im}~\chi_i & {\rm Re}~\chi_i \cr
0 & 0 \cr
 {\rm Re}~\phi_i & {\rm Im}~\phi_i \cr
- {\rm Im}~\phi_i & {\rm Re}~\phi_i \cr
} \right).
}

\newsec{Holographic renormalization group flows}

\subsec{Some simple tests}

There are several obvious consistency checks upon
our putative dual of a subsector of the quiver theories.
First is the fact that the $SU(1,n)$ symmetry discovered
above is essential from the field theory perspective.
Recall that the $SU(1,n)$ appeared as the commutant
of the $SU(2) \times U(1)$ gauge symmetry, which means
that all the supergravity scalars in the coset:
\eqn\cstagain{\cT ~=~ {SU(1,n) \over U(1)\times SU(n) } \,, }
are precisely the ${\bf 1}_0$ of the R-symmetry.  As we noted earlier,
these are the  
duals of the complex gauge coupling constants, $\tau_i$, on the nodes
of the quiver theory.  The fact that the supergravity potential has an
$SU(1,n)$  
invariance means that these coupling constants are freely chooseable
parameters in the  
theory, as they must be from the brane perspective. This fact is the
generalization to  
the quiver theory of the $SU(1,1)$ symmetry of the IIB theory and its
relationship to  
the complex gauge coupling in the $\cN=4$ Yang-Mills theory.

One can also check that the results of \DallAgataVB\ and
of the previous sections reproduce the solutions of
\refs{\KPW,\FGPWa}.  To be more precise, in \KPW\ 
a general $SO(5,2)$ matrix was constructed (up to 
symmetries of the potential), and the $\cN=8$ supergravity
potential was then constructed for this matrix.  We have taken
the same $SO(5,2)$ matrix and have explicitly verified 
that \potential\ and \STUVtens, with the  choices of 
${\Lambda^{N}}_{M}$  and ${f_{IK}}^{J}$ in section 4,
exactly reproduces the scalar potential of \KPW.

The flow studied in \FGPWa\  used two parameters.  The first, denoted
$\rho = e^\alpha$,  
was the $SO(1,1)$ group element, while the second was a particularly
simple $SO(5,2)$  
matrix: 
\eqn\flowMat{ L ~=~\exp(\cX) \, \ \ {\rm where} \  \ \
 \cX_{1,6} = \cX_{6,1} =\chi \,, \ \ 
 \cX_{2,7} = \cX_{7,2} =\chi\,,}
with all other matrix elements equal to zero (this corresponds to
${\rm Re}~\chi_1=\chi$ and all other modes turned off in~\submatrix ).   
The $\cN=1$ supersymmetric flow is characterized by
a superpotential, $W$, that is related to the scalar potential
via:
\eqn\VfromW{\cV ~=~ {g^2 \over 8}~\sum_{j = 1}^2 
~\Big| {\del W\over \del \varphi_j} \Big|^2 
~-~ {g^2 \over 3}~\big|W \big|^2 \,,}
with $\varphi_1 =\chi$  and $\varphi_2 = \sqrt{6}~\alpha$.
The scalars, $\varphi_j$,  are defined so as to have canonically 
normalized  kinetic terms.  The superpotential was computed to be:
\eqn\Wreduced{W~=~ {1 \over 4 \rho^2}~ \Big[\cosh(2 \varphi_1)~(
\rho^{6}~-~ 2)~ - ( 3\rho^{6} ~+~ 2 ) \Big] \ ,} 
and the equations of motion for the flow are:
\eqn\steepdesc{{d \varphi_j \over d r} ~=~{g \over 2}~{\del W \over
\del \varphi_j} \ , \qquad  
{d\, A \over d\,r}~=~ - {g \over 3}~W \ ,}
where $A(r)$ is the cosmological function in the five-dimensional
metric: 
\eqn\RGFmetric{ds^2 =   dr^2 ~+~ e^{2 A(r)} \eta_{\mu\nu} dx^\mu dx^\nu\ .}
 The superpotential gives rise to two supersymmetric ground states: 
The maximally symmetric, and maximally supersymmetric one
 with $\varphi_j=0$, and the other with:
\eqn\susyGS{\varphi_1 ~=~ \pm \coeff{1}{2}~\log(3) \ ,  
 \quad \varphi_2  \equiv \sqrt{6}~ \alpha ~=~\coeff{1}{\sqrt{6}} \log(2)   \ .}
This second point has $\cN=2$ supersymmetry in the bulk, and
on the brane it corresponds to the $\cN=1$ superconformal point of
Leigh and Strassler \refs{\RGLMJS,\FGPWa}.     The flow of
interest is the steepest descent from the $\varphi_j=0$
critical point to either of the non-trivial points \susyGS.  The 
cosmological  constants at the two critical points are, respectively 
$-{3 \over 4} g^2$ and $-{1\over 3} 2^{4/3}  g^2$, which leads to the
result  that ${c_{IR} \over c_{UV}} = {27 \over 32}$.

As we described earlier,  the  scalars $\chi$ and $\alpha$ are
respectively dual to the following operators
in the $\cN=4$ Yang-Mills theory:
\eqn\twoflowops{   \cO_f \eql    
{\rm Tr} \big( \lambda^3\lambda^3 \big) \,, 
\qquad \cO_b \eql 
- \,\sum_{j=1}^4\,  {\rm Tr} \big( X^jX^j \big)  \,
+\, 2\sum_{j=5}^6 {\rm Tr}\big( X^{j} X^{j})   \,,}
and the flow corresponds to turning on a mass for a single
chiral multiplet.

In the $\cN=8$ supergravity theory the superpotential can be
read off from the eigenvalues of a scalar matrix in the 
gravitino variation.  We have indeed confirmed that
the same results are obtained from the gravitino variation in 
 \DallAgataVB\ using the scalars $\alpha$ and $\chi$ defined 
above.

\subsec{New flows from old}
 
The unbroken global symmetry $SU(1,n)$ in the
massless sector of the $\BZ_{n}$ orbifolds has profound
consequences for the holographic descriptions of renormalization
group flows in the corresponding quiver gauge theories.

The $SO(5,2)$ matrix defined in \flowMat\  can be embedded directly
into $SO(5,2n+1)$ or  
$SO(5,2n+3)$ resulting in a particular flow solution for the 
quiver gauge theory.  This flow
corresponds to the deformation of the quiver theory by a single
adjoint mass term which 
breaks $\CN=2$ to $\CN=1$ supersymmetry on the brane. This
embedding is guaranteed to be consistent   
because the other matter multiplets can be
consistently truncated out, using an $SU(n-1)\times \BZ_{2}$ symmetry
for $\BZ_{n>2}$   or a $U(1)\times SU(2)_L$ symmetry
for $\BZ_{2}$.  Now consider 
the action of the $SU(n)$ ``flavor symmetry.'' This can map the matrix
elements of $\cX$  
onto any matrix with
\eqn\flowMat{ \cX_{1,2j+4} = \cX_{2j+4,1} =\chi_j \,, \ \ 
 \cX_{2,2j+5} = \cX_{2j+5,2} =\chi_j\,, \qquad j = 1,\dots, n  \,.}
The parameters $\chi_j$ are the supergravity scalars that 
represent the couplings $m_j$ in \Genopdef.  

The potential and superpotential are invariant under the 
$SU(n)$ and so we conclude that the critical point
described above extends to a critical surface swept out
by the $SU(n)$ action.  Similarly, the flow of  \FGPWa\ 
becomes a continuous family of flows to this critical surface,
with the end-point determined by the initial velocities
$\chi_j$, or $m_j$.  
We therefore conclude that, at least
for large $N$, the quiver theories have a continuous family
of $\cN=1$ supersymmetric fixed points with ${c_{IR} \over 
c_{UV}} = {27 \over 32}$, and these are all swept out 
by the continuous action of $SU(n)$.  In particular, the
solutions of  \IKEW, and its generalizations in \GubserIA,
and the solution of \FGPWa\ are simply $SU(n)$ images of
one another, and merely represent isolated solutions
in a continuum family. To be precise, for $n=2$, the deformation 
by~\Aoneopdef\ corresponds to the flow with
${\rm Re}~\chi_1=-{\rm Re}~\chi_2=\chi/2$, while that of~\Aoneuntwist\
has\foot{More generally, the
untwisted deformation corresponds to ${\rm Re}~\chi_i={\chi \over n}$
for all $i$.}   ${\rm Re}~\chi_1={\rm Re}~\chi_2=\chi/2$.

It is interesting to note that the gauged $\cN=4$ supergravity
theory contains scalars, $\chi_j$, that are dual to fermion 
masses in each node of the quiver theory.  On the other
hand, the only scalar mass term in the $\cN=2$ vector multiplet
on the brane that is ``resolvable'' 
within this gauged supergravity is \dildual: the sum over all the 
nodes of the quiver.   In any supersymmetric flow, the scalar
masses are determined by the fermion masses, and so the scalar mass 
data is inessential to the study of the generalizations of
the  ``Leigh-Strassler'' flows.  It was  however noted in \FGPWa\ that
one could apparently give independent initial velocities
to the supergravity scalars that are dual to the fermion
and scalar bilinears on the brane, and still preserve 
supersymmetry.  These more general flows could then be
interpreted as turning on a mass for a chiral multiplet,
and flowing out along the Higgs branch of the remaining
massless multiplets.  The gauged supergravity considered
here still captures these flows for the ``sum over nodes'' on
the quiver, but apparently not within each individual node of the
quiver separately.

\newsec{Discussion}

Motivated by the many successes of gauged supergravity theory in
studying holographic RG flows, we have made some progress in extending
the toolkit to spacetime orbifolds and their field theory duals. The
applicability of our approach to the massless string modes in the 
untwisted sector is guaranteed by virtue of the consistent truncation
to $\Gamma$--invariant subsectors of the maximal gauged supergravity
theory in five dimensions. However, we have also described the way in
which one can incorporate the twisted sector degrees of freedom. Here
we have passed beyond the safe haven of consistent truncations and we
cannot, {\it a priori}, guarantee that the solutions we obtain in our
five-dimensional theory always 
correspond to  ten-dimensional solutions of the full IIB theory. 

\subsec{Features   of the gauged supergravity approach}

We have presented several encouraging facets of our approach which
motivate our trust in it. First of all, the structure of the 
$\cN=4$, five-dimensional gauged supergravity theories found
in~\DallAgataVB\ is 
completely determined by the embedding of the gauge group in the
scalar coset. We found that the IIB massless modes had a particular
charge structure under the $SU(2)_R\times U(1)$ group that was to be
gauged, and so the gauged supergravity theory that is dual to quiver
gauge theories is unique up to field re-definitions.

We have also used our approach to examine the family of RG flows
generated by  
perturbations of the quiver theory by masses for the adjoint scalar
fields. This yielded  
a picture of a $\BP^{n-1}$ critical surface of superconformal fixed
points, which was  
precisely in line with our field theory analysis in section~2. The
gauged supergravity  
results also indicate that the value of the central charge on these
surfaces is ${c_{IR} \over c_{UV}} = {27 \over 32}$, which is
consistent with the calculations  of~\refs{\GubserVD,\BergmanQI},  
as well as those of section~2. 

In several of these RG flows, ten-dimensional solutions are
known. In~\IKEW, 
it was argued that IIB on the conifold was the theory that described
the IR fixed point of the flow generated by
the deformation of the $n=2$ quiver theory by the twisted
operator~\Aoneopdef. From the orbifold connection between $\cN=4$
Yang-Mills and the $\cN=2$ quiver theories, it is easy to see that the
ten-dimensional lift of the flow generated by the untwisted operator
\eqn\untwistedop{
\CW_u = {m\over 2}\, \sum_{i=1}^n\,  \int d^2\theta\, \Phi_i^2,
}
is given by a $\BZ_n$--orbifold of the ``Leigh--Strassler'' solution
found 
in~\refs{\KPNWa,\KPNWc}\foot{To be precise, this $\BZ_n$ acts as an
identification on the Hopf angle, $\alpha_3$,  of the $SU(2)$
submanifold 
defined in equation~(3.7) of \KPNWc. This means that $SU(2)$ is 
replaced by the Lens space $S^3/\BZ_n$. The rest of the
ten-dimensional solution, 
including the 2-form and 5-form backgrounds, is
invariant under the $\BZ_n$ action, and so is unmodified.}. Manifolds
which provide a basis for the ten-dimensional description 
of the general flows were described in~\GubserIA\ and these were
used in the computation of the central charges~\BergmanQI\ that our
results agree with. 

The fact that there are known ten-dimensional solutions
that coincide with individual points in the families of
five-dimensional solutions, 
and that the results of~\GubserIA\ suggest that these ten-dimensional
solutions also form families with the same symmetries and
supersymmetry, is consistent with our conclusions here.  

It would certainly be more
satisfying (as well as extremely useful in practice) if we could
directly extend the technology of uplifts of five-dimensional solutions to
ten-dimensional IIB solutions   (as in, for example, 
\refs{\KPNWa,\KPNWc, \KhavaevYG}) to the solutions of $\cN=4$
gauged supergravity theories. An important part of this program involves
the correct approach to include the geometric data of topology
change due to the partial resolutions of the orbifold singularity in
the general case. This is currently under investigation.

We have reproduced the branch of moduli space which corresponds to the
adjoint masses, but we have had less success in reproducing the moduli
space, $\CM_g^{(n)}$ of gauge couplings found
in~\refs{\WittenSC,\DoreyQJ}.  The duality group can be expressed as a
certain extension of $SL(2,\BZ)$ by a braid group\foot{We thank E.\
Witten for discussions on these points.}. The resulting discrete group
appears to be too large to be embedded in the $SU(1,n)$ symmetry group
found here.  Presumably the difference lies in the large $N$ limit
that is inherent in the supergravity approach.  This limit may, on
the one hand, collapse or trivialize some of the quantum duality
symmetry, while on the other hand promote another part of it
to a continuum symmetry.  It is worth recalling that for $n=1$ the
$SU(1,1)$ symmetry of the IIB supergravity is reduced to  
$SL(2,\BZ)$ in the string theory.   It would
be interesting to determine what happens to the
$SU(1,n)$ symmetry of the supergravity theory considered here.
A physically natural guess might involve a symmetry that leaves
the hyperplanes, $m_i =0$ fixed, and so might involve a
combination of the Weyl group of $SU(n)$ and the $SL(2,\BZ)$
action inherited from the IIB string.

\subsec{Additional Interesting Flows}

We have studied one class of flows, but there are other interesting
deformations that can be studied within the $\cN=4$ gauged
supergravity. For example, one component of the vector multiplet
in the untwisted sector is dual to a $\BZ_n$-invariant hypermultiplet
mass term, $\sum a_i b_i$, which preserves $\cN=2$ supersymmetry on the
D3-brane. This 
could be used to probe the vacuum of the quiver theory, as well as of
other fixed points with less supersymmetry. 

More interesting flows can be generated by other initial conditions
for the scalars $\phi_i$ and $\chi_i$. Since the $\phi_i$
that are dual to coupling constants in the field theory, the coupling
constants can run along these flows. It would be interesting to see if
there are flows, perhaps incorporating additional domain wall
configurations in five dimensions, which  result in $H_{RR}$ flux in
ten-dimensions that could be 
interpreted as fractional brane configurations. 

There are some flows which might be outside of the scope of
the holographic approach within IIB supergravity alone.   
For example, there are non-$\BZ_n$-invariant chiral primary
operators  $a_i b_i$ (no sums on $i$) that do not
have dual fields in the twisted sector discussed in section~3.  This
is consistent with the notion that the non-trivial RR flux has frozen
out some moduli (like the FI-terms for the gauge theory).  

We also note some features of the operator spectrum of the
$\CN=2$ algebra in 
four-dimensions. We are able to find the duals of individual mass
terms for the 
fermions in the vector and hypermultiplets. Indeed the fermion masses
in the vector multiplets
are what we used to generate the $\CN=1$ flows that we studied. On the
other hand, the individual {\it boson} mass terms like $|a_i|^2$,
$|b_i|^2$, and $|\phi_i|^2$ are apparently non-chiral. This can be
most directly seen as a consequence of the fact that, in the projection
from the $\CN=4$ SYM theory, they correspond to components of bilinear 
operators which are not chiral.  Instead we see only linear combinations of
the 
operators~\dildual\ and $\sum_i^n(|a_i|^2-|b_i|^2)$, which are related
to  components of chiral $\CN=4$ operators.

\subsec{Surprising Implications of Gauged Supergravity Symmetries}

Perhaps the most intriguing result of our analysis is the $SU(1,n)$
symmetry on the tensor multiplet sector that we found in section~4.  
One may think of this symmetry group is
in terms of the ``bonus symmetries'' \IntriligatorIG\ that arise from
enhancements of discrete symmetries into continuous ones in passing
from a  string theory to its supergravity limit. One example is the
promotion 
of the $SL(2,\BZ)$ S-duality group to $SU(1,1)$. 
In the orbifolds we
have studied here, an additional discrete symmetry is the $\BZ_n$
quantum symmetry, which leads to a $SL(2,\BZ) \times \BZ_n$ group
which is being promoted to $SU(1,n)$.  

One can gain a different perspective on this by noting 
that  the six-dimensional theory of IIB at $\BC^2/\BZ_n$ has a 
scalar coset $SO(5,n+1)/[SO(5)\times SO(n+1)]$. The full duality group
of the IIB 
string is obtained from counting the tensor multiplets and is
$SO(2,n+1;\BZ)$. In the near horizon limit, we saw 
that this space is enlarged to form the coset of five-dimensional
scalars. Apparently the $SU(1,n)$ symmetry is non-trivially related 
to the residual  IIB duality group in the near-horizon background. 
This would suggest another obvious guess of $SU(1,n; \BZ)$ for
the quantum duality symmetry.

This discussion illuminates the
extension of our results to the $D$ and $E$ series. In six dimensions,
we will find scalar cosets $SO(5,r+2)/[SO(5)\times SO(r+2)]$ and
duality group $SO(2,r+2;\BZ)$, where $r$ is the order of the orbifold
group. The KK-reduction of the $r$
twisted-sector tensor multiplets will again lead to $2r$ charged
five-dimensional tensor multiplets with the same $SU(2)_R\times U(1)$
charges 
of ${\bf 1}_{\pm 2}$ as we find in the untwisted sector. The
linearized fields will have kinetic terms which are 
weighted by the relevant Cartan/intersection matrix, as in~\normkin,
so we will make the same type of basis change in going to the
five-dimensional fields as we described in section~4.5.  Accounting
for the single neutral vector multiplet we obtain in the untwisted
sector, we find a scalar coset $SO(5,2(r+1)+1)/[SO(5)\times
SO(2(r+1)+1)]$. The gauging proceeds as in the $A$-series case and we
find an analogous $SU(1,r+1)$ symmetry.

From the perspective of supergravity we might have
anticipated the $SU(1,n)$ symmetry  from the outset.  The fact
that the quiver theories are superconformal with $n$ complex
couplings, $\tau_j$ that can be chosen freely at the UV fixed point
means that the corresponding supergravity must have $n$ complex scalars
that represent flat directions in the supergravity potential.   The
presence 
of a $\BZ_n$ quantum symmetry that cycles the nodes of
the quiver, and hence the gauge couplings, combined with the 
original $SU(1,1)/U(1)$ coset of the IIB theory points 
rather directly at $SU(1,n)$.

There are also consequences of the $SU(n)\subset U(1,n)$ symmetry
for the structure of ten-dimensional solutions. This symmetry relates fields in
the untwisted sector of the IIB orbifold theory with those in the
twisted sector. In fact, in five-dimensional terms, it puts these
fields on the same footing. On the other hand, in terms of the ten-dimensional 
geometry, turning on these fields does very different things. For example, we have
already described the fixed point geometry dual to the fixed point
generated by the untwisted deformations~\untwistedop\ in terms of an
orbifold of the ten-dimensional solution in~\KPNWc. This involves D3-branes
transverse to a space with a $\BZ_n$ orbifold singularity and certain
backgrounds for the self-dual 5-form and the 2-form $B$-fields. We
argued that, for $n=2$, this fixed point is related by $SU(2)$
to the fixed point which is generated by the twisted
deformation~\Aoneopdef. The ten-dimensional solution which governs the
fixed point 
in this case~\IKEW\ involves D3-branes transverse to the conifold (the
original $\BZ_2$ orbifold singularity has been removed),
with an appropriate 5-form field strength turned on, but {\it no}
$B$-fields.  For $n>2$, the analogous fixed points are dual to
cones over  
the level surfaces of the generalized conifolds constructed in~\GubserVD. 
It is also expected that the IIB solution in these cases does not involve
$B$-fields.  

Finally, we believe 
that this $SU(1,n)$ symmetry can also be used to make further
statements about 
these fixed point theories. For example, we expect that the coupling
constant dependence of correlation functions at large~$N$ should be
constrained to $SU(n)$ covariant expressions. 

\bigskip
\leftline{\bf Acknowledgements} The work of R.C. and N.W. was
supported in part by funds  
provided by the DOE under grant number DE-FG03-84ER-40168. The work of
M.G.\ was  
supported in part by the National Science Foundation under Grant
Number PHYS-0099548  
and  he would like to thank the CITUSC Center for Theoretical Physics for its 
hospitality where part  of this work was done. The work of M.Z. was
supported in part by  
the German Science Foundation (DFG) within the ``Schwerpunktprogramm
1096". M.Z would  
like to thank the CIT-USC  Center for Theoretical Physics and the
Physics Department of  
the Pennsylvania State University for their
 hospitality while
part  of this work was done.
R.C. and N.W.\ would like to
thank D.\ Berenstein, P.\ Berglund, J.\ Gomis, A.\ Hanany, C.\ Herzog,
I.\ Klebanov,
K.\ Pilch,  C.\ R\"omelsberger, and E.\ Witten for discussions. M.Z.\ would like
to thank C.\  
Herrmann for discussions.

\vfill
\eject 
\appendix{A}{Analysis of the scalar manifold}

In this appendix we will determine the total number of independent
scalar fields of the  
potential of gauged $\CN=4$ supergravity, with the gauge group
$SU(2)_R\times U(1)$,  
coupled to $2n$ tensor multiplets with or without  an extra spectator
vector multiplet   
that could drive renormalization group flows. The scalar fields of
these  $\CN=4$   
supergravity theories
 parameterize the coset space
$$
\CS=SO(1,1)\times {SO(5,p) \over SO(5)\times SO(p)},
$$
where $p=2n $ for the coupling of $2n$  tensor multiplets and 
$p=2n+1$ when we have an additional spectator vector multiplet.

For $p=2n$ or $p=2n+1$ the relevant gauged supergravity has the
residual symmetry  
$SU(2)_R \times U(1) \times SU(1,n)\times SO(1,1)$ \foot{ As discussed
earlier we are  
considering the case when all the pairs of tensor fields carry equal
and opposite  
charges with respect to $U(1)$.}. Let us first consider the case
$p=2n$. Then the  
$10n$ scalar fields of the coset $SO(5,2n)/[SO(5)\times SO(2n)]$ decompose as
 $(5,2n)$ under the subgroup
$SO(5)\times SO(2n)$. With respect to the $U(n)$ subgroup of
$SU(1,n)\subset SO(2,2n)$ each of the five sets of $2n$
scalars in the fundamental representation of $SO(2n)$ decomposes
as $n \oplus \bar{n}$. We shall label them as 
$$\eqalign{
& Z^i_A \simeq  (5,n) \in  SO(5)\times U(n), \cr & Z^{iA} \simeq (5,
\bar{n}) \in  
SO(5)\times U(n), } $$ where $ i,j,\ldots=1,\ldots,5 $ and
$A=1,\ldots,n$. Consider now  
the general case when $n \geq 4$. By using the noncompact symmetries
of the residual  
$SU(1,n)$ symmetry we can gauge away one set of $n + \bar{n}$ scalars,
namely a linear  
combination of $Z^4_A$ and $Z^5_A$ , leaving us with four sets of
$n+\bar{n}$ scalars.   
Now by the local $U(n)$ symmetry we can rotate the first set $Z^1_A$
to the vector  
$(R^1_1,0,...,0)$ where $R^1_1$ is real. Using the little group
$U(n-1)$ of this vector  
we can rotate the second vector $Z^2_A$ into a vector of the form 
$(C^2_1,R^2_2,0,...,0)$ where $C^2_1(R^2_2)$ is complex
(real). Similarly $Z^3_A$ can be  
brought to the form $(C^3_1,C^3_2, R^3_3,0,..,0)$ by a $U(n-2)$
rotation. The remaining  
linear combination of $Z^4_A$ and $Z^5_A$ can similarly be brought to the form 
$(C^4_1,C^4_2,C^4_3,R^4_4,0,..,0)$. All this leaves us with 16 real
scalars. Using the  
$SU(2)_R\times U(1)$ gauge symmetry we can gauge away 4 of the 16
scalars leaving us  
with 12 scalars plus the $SO(1,1)$ dilaton.

For $n <4$ the above analysis needs to be slightly modified.
For $n=1$ we are left with  1 real and 3 complex scalars from the
coset $SO(5,2)/[SO(5)\times SO(2)]$  after gauge fixing the $SU(1,1)$
symmetry fully. 
Further fixing of the $SU(2)_R\times U(1)$ gauge symmetry leaves
us with 3 real scalars plus the $SO(1,1)$ dilaton, thus agreeing
with the analysis of \KPW. For $n=2$ we get $5\times 2+2=12$
 scalars surviving  
from the coset $SO(5,4)/[SO(5)\times SO(4)]$. after utilizing the
$SU(1,2)$ symmetry.  
Then fully gauge-fixing the $SU(2)_R\times U(1)$ symmetry we are left
with 8 scalars  
plus the $SO(1,1)$ dilaton. Similar analysis yields 11 real scalars
plus the $SO(1,1)$  
dilaton for $n=3$  after we fix all the gauges.

The  above results  generalize trivially to the cases when there is an
additional spectator
vector multiplet  present and the scalar
manifold is $SO(5,2n+1)/SO(5)\times SO(2n+1)$  since the extra five scalars
sitting in the coset $SO(5,2n+1)/SO(5)\times SO(2n+1)$ are all singlets of the
residual $SU(1,n)$ symmetry of the gauged supergravity describing
the coupling of $2$n tensor multiplets and spectator  vector multiplet
to $\cN=4$ supergravity. Thus the net number of scalar fields left
after gauge fixing  all the symmetries is 5 more than the
corresponding cases without the extra vector multiplet.

The fact that the number of independent moduli does not grow with
increasing $n>4$ may at first sight appear surprising. However we
should keep in mind that the holographic theories we are trying to
understand all arise from orbifolding of the five sphere and the
total moduli available for the renormalization group flows should
not increase by the orbifolding procedure. The scalar manifold of
the $\cN=8$ supergravity in $d=5$ is $E_{6(6)}/USp(8)$ corresponding
to 42 scalars. The $S^5$ compactification of IIB superstring leads
to gauged $\cN=8$ supergravity with the gauge group $SU(4)$ and an
additional $SU(1,1)$ symmetry \GRW. Gauge fixing of all the
symmetries of this theory leaves us with 42-18=24 scalar fields
that could drive renormalization group
flows.

\listrefs
\vfill
\eject
\end